\documentclass[10pt,aps,pra,twocolumn,superscriptaddress,floatfix,nofootinbib]{revtex4-2}

\usepackage{amsmath, amsfonts,mathrsfs, amssymb,mathtools, commath}
\usepackage{bm,bbm, braket, accents}
\usepackage{graphicx}   
\usepackage[usenames,dvipsnames]{color}
\usepackage{algpseudocode,algorithm,algorithmicx}

\makeatletter
\def\@bibdataout@aps{%
 \immediate\write\@bibdataout{%
  @CONTROL{%
   apsrev41Control,author="08",editor="1",pages="0",title="0",year="1",eprint="1"%
  }%
 }%
 \if@filesw
  \immediate\write\@auxout{\string\citation{apsrev41Control}}%
 \fi
}%
\makeatother 

\newcommand{\ketbra}[2]{\left\vert{#1}\right\rangle\!\left\langle{#2}\right\vert}
\newcommand\Tr{\mathrm{Tr}}
\newcommand{\iprod}[2]{\left\langle{#1}\middle\vert{#2}\right\rangle}
\newcommand{\oprod}[2]{\left\vert{#1}\middle\rangle\!\middle\langle{#2}\right\vert}

\usepackage[breaklinks=true]{hyperref}
\hypersetup{
  colorlinks   = true, 
  urlcolor     = blue, 
  linkcolor    = blue, 
  citecolor   = red 
}

\usepackage[dvipsnames]{xcolor}

\usepackage[deletedmarkup=sout, authormarkup=superscript]{changes}
\definechangesauthor[color=Red]{ToDo}
\definechangesauthor[color=Plum]{MN}
\definechangesauthor[color=ForestGreen]{JC}

\newtheorem{rmk}{Remark}

\usepackage[capitalise]{cleveref} 
\crefformat{equation}{Eq.~(#2#1#3)} 
\crefformat{section}{Sec.~#2#1#3} 
\Crefformat{equation}{Equation~(#2#1#3)}
\crefformat{figure}{Fig.~#2#1#3}
\crefrangeformat{equation}{Eqs.~#3(#1)#4--#5(#2)#6}
\Crefformat{section}{Section~#2#1#3}

\begin{document}

\title{Solving quantum optimal control problems using projection operator-based Newton steps}

\author{Jieqiu Shao}
\affiliation{College of Engineering and Applied Science, University of Colorado Boulder, Boulder, Colorado 80309, USA}

\author{Mantas Naris}
\affiliation{College of Engineering and Applied Science, University of Colorado Boulder, Boulder, Colorado 80309, USA}

\author{John Hauser}
\affiliation{College of Engineering and Applied Science, University of Colorado Boulder, Boulder, Colorado 80309, USA}

\author{Marco M. Nicotra}
\affiliation{College of Engineering and Applied Science, University of Colorado Boulder, Boulder, Colorado 80309, USA}

\date{\today}

\begin{abstract}
The Quantum PRojection Operator-based Newton method for Trajectory Optimization, a.k.a. Q-PRONTO, is a numerical method for solving quantum optimal control problems. This paper significantly improves prior versions of the quantum projection operator by introducing a regulator that stabilizes the solution estimate at every iteration. This modification is shown to not only improve the convergence rate of the algorithm, but also steer the solver towards better local minima compared to the un-regulated case. Numerical examples showcase how Q-PRONTO can be used to solve multi-input quantum optimal control problems featuring time-varying costs and undesirable populations that ought to be avoided during the transient. 
\end{abstract}

\maketitle

\section{Introduction}\label{sec:intro}
Quantum optimal control is an enabling technology that allows quantum systems to operate at their fullest potential   \cite{brif2010control,dong2010quantum,altafini2012modeling,glaser2015training}. Model-based solvers predominantly rely on the GRadient Ascent Pulse Engineering algorithm (GRAPE) \cite{khaneja2001time,khaneja2005optimal} or the Krotov method \cite{sklarz2002loading,reich2012monotonically}, both of which feature a linear convergence rate, or superlinear in the case of gradient acceleration schemes \cite{de2011second,eitan2011optimal}. In prior work, the authors introduced Q-PRONTO \cite{jay2022}: a Newton-based  quantum optimal control solver that can achieve quadratic convergence.\smallskip

Although Q-PRONTO is a quantum adaptation of the general-purpose Projection Operator-based Newton method for Trajectory Optimization \cite{hauser2002projection,john2003}, the prior version \cite{jay2022} featured a simplified formulation that did not include a regulator. The main contribution of this paper is to augment Q-PRONTO's projection operator with a regulator term. This improves the numerical stability of the algorithm, thereby boosting its convergence rate by enabling it to take larger steps at each iteration.
\smallskip

In addition to improving the convergence rate of the Q-PRONTO algorithm by introducing the regulator, the paper extends the range of quantum optimal control problems it can address by detailing how to i) perform both phase-sensitive and phase-insensitive state transformations, ii) penalize undesirable states during he transient behavior, iii) handle multiple inputs,iv) use idealized behaviors as an initial guess for the solver, , and v) design quantum gates.\smallskip

The paper is structured as follows: Section \ref{sec:prob_statement} introduces the quantum control problem and shows how to reformulate it as a real-valued optimal control problem. Section \ref{sec:PRONTO} redevelops the Q-PRONTO formulation in the presence of a possibly non-zero feedback regulator. Section \ref{sec:regulator} details how to design the regulator depending on the control objectives. Section \ref{sec:NumEx} features three numerical examples: the first illustrates the benefits of using a non-zero feedback gain, the second highlights the novel capabilities of Q-PRONTO, and the third show how to design a quantum gate.

\section{Problem Statement}\label{sec:prob_statement}
The objective of this paper is to develop a numerically efficient algorithm for solving quantum optimal control problems \cite{brif2010control} in the form 
\begin{subequations}\label{eq:ocp_Quantum}
\begin{eqnarray}
\displaystyle \min ~~&&\displaystyle \mathcal J\bigl(\ket{\psi(T)}\bigr)+\int_0^T \ell\bigl(\ket{\psi(t)},u(t),t\bigr)dt \label{eq:ocp_Qcost}\\
\textrm{s.t.}~~&& 
\mathrm{i}\hbar\ket{\dot\psi(t)} = \mathcal H\bigl(u(t)\bigr)\ket{\psi(t)},\qquad \ket{\psi(0)}=\ket{\psi_0},~~~\label{eq:Schrodinger}
\end{eqnarray}
\end{subequations}
where \eqref{eq:Schrodinger} is the Shr\"odinger equation of a closed quantum system governed by the control Hamiltonian $\mathcal{H}(u)$ and 
\eqref{eq:ocp_Qcost} is a cost function that encapsulates the control objectives. Note that $\ket{\psi(t)}$ can be interpreted as either a wavefunction or a vector of wavefunctions. As detailed in Section \ref{ssec:gates}, the latter is useful for quantum gate design. Each component is characterized as follows.

\subsection{Schr\"odinger Equation}
Given the unit wavefunction $\ket\psi\in\mathbb C^n$ and the control input vector $u\in\mathbb R^m$, the Shr\"odinger equation \eqref{eq:Schrodinger} is governed by the control Hamiltonian
\begin{equation}
    \mathcal H(u) = \mathcal H_0 + \sum_{j=1}^m\mathcal H_j\, f_j(u_j),
\end{equation}
where $\mathcal H_0$ describes the free evolution of the system, $\mathcal H_j$ describes the evolution induced by the $j$-th control input, and $f_j:\mathbb R \to \mathbb R$ are class $\mathcal{C}^2$ functions that capture possible nonlinearities. To ensure that $\|\!\ket{\psi(t)}\!\|$ is constant, the matrices $\mathcal H_0$, $\mathcal H_j$ are assumed to be Hermitian.

\subsection{Cost Function}
The cost function \eqref{eq:ocp_Qcost} is comprised of a terminal cost $\mathcal{J}(\ket\psi)$ and an incremental cost $\ell(\ket\psi,u,t)$. The former penalizes deviations between a given objective and the final wavefunction $\ket{\psi(T)}$, whereas the latter penalizes undesirable behaviors during the time window $t\in[0,T]$. To ensure the existence and uniqueness of the solution, it is assumed that $\mathcal{J}:\mathbb C^n\to\mathbb R$ is a class $\mathcal{C}^2$ convex function and
$\ell:\mathbb C^n\times \mathbb R^m\times[0,T]\to\mathbb R$ is a class $\mathcal{C}^2$ function that is convex in $\ket\psi$ and \emph{strongly} convex in $u$ for all $t\in[0,T]$. \smallskip

For the reader's convenience, the following paragraphs provide suitable examples of cost functions when $\ket{\psi}$ is a single wavefunction. Although these cost functions stem from the ones featured in \cite{brif2010control}, they have been manipulated to satisfy the convexity assumption.  \medskip

\noindent\textbf{Terminal Cost}\\
The terminal cost typically captures the main control objective of the quantum optimal control problem. Indeed, \eqref{eq:ocp_Quantum} aims to steer the wavefunction from an initial condition $\ket{\psi_0}$ to a final condition $\ket{\psi_T}$ that, ideally, satisfies $\mathcal J(\ket{\psi_T})=0$. The choice of $\mathcal{J}(\cdot)$ depends on the global phase that is allowed between the target $\ket{\psi_T}$ and the final wavefunction $\ket{\psi(T)}$.\medskip

\noindent\emph{\underline{Zero Phase Error:}} If the global phase error between $\ket{\psi_T}$ and $\ket{\psi(T)}$ matters, one would typically choose
\begin{equation}\label{eq:ZeroPhaseError}
    \mathcal J(\ket{\psi(T)})=1-\mathrm{Re}\left(\iprod{\psi(T)}{\psi_T}\right).
\end{equation}
Unfortunately, this function is not convex. However, it can be rewritten as a convex function by noting that, given two unit wavefuctions $\ket\psi$, $\ket\phi$, it is possible to write $\iprod\psi\psi+\iprod\phi\phi=2$ and $\iprod{\psi}{\phi}+\iprod{\phi}{\psi}=2\,\mathrm{Re}(\iprod{\psi}{\phi})$. Based on these properties, \eqref{eq:ZeroPhaseError} can be reformulated as
\begin{equation}
    \mathcal J(\ket{\psi(T)})=\tfrac12\left(\bra{\psi(T)}-\bra{\psi_T}\right)\left(\ket{\psi(T)}-\ket{\psi_T}\right),
\end{equation}
which is a class $\mathcal C^2$ convex function of $\ket{\psi(T)}$.
\medskip

\noindent\emph{\underline{Arbitrary Phase:}} If the global phase error between $\ket{\psi_T}$ and $\ket{\psi(T)}$ does not matter, a suitable choice for the cost function would be the Hilbert-Schmidt distance
\begin{equation}\label{eq:GlobalPhaseError}
    \mathcal J(\ket{\psi(T)})=1-\left|\iprod{\psi(T)}{\psi_T}\right|^2.
\end{equation}
Although this function is also not convex, it can be rewritten as a convex function by noting that $\iprod{\psi}{\psi}=1$ and $|\!\iprod{\psi}{\phi}\!|^2=\iprod{\psi}{\phi}\iprod{\phi}{\psi}$. Based on these properties, \eqref{eq:GlobalPhaseError} can be rewritten (up to a factor $2$) as the class $\mathcal C^2$ convex function 
\begin{equation}
    \mathcal J(\ket{\psi(T)})=\tfrac12\bra{\psi(T)} \Gamma_T \ket{\psi(T)},
\end{equation}
where $\Gamma_T=(I\!-\!\oprod{\psi_T}{\psi_T})$ is a positive definite matrix.
\medskip

\noindent\textbf{Incremental Cost}\\
The incremental cost is an auxiliary term that ensures that the solution is ``well behaved'' during the time window $t\in[0,T]$. Unless there are specific reasons to penalize the cross-correlation between $\ket\psi$ and $u$, the most common incremental cost is
\begin{equation}\label{eq:qincrimental_cost}
    \ell (\ket{\psi}\!,u,t\big)= \ell_\psi(\ket{\psi}\!,t) + \ell_u(u,t),
\end{equation}
with $\ell_\psi\!:\mathbb C^n\times[0,T]\!\to\!\mathbb R$ convex in $\ket\psi$ for all $t\in[0,T]$ and $\ell_u\!:\mathbb R^m\times[0,T]\!\to\!\mathbb R$ strongly convex in $u$ for all $t\in[0,T]$ to ensure that $u(t)$ does not grow unbounded.\medskip

\noindent\emph{\underline{Control Effort:}} A typical choice for the incremental input cost is the class $\mathcal C^2$ strongly convex function
\begin{equation}
    \ell_u(u,t)=\tfrac12u^T \mathcal R(t)\, u,
\end{equation}
where $\mathcal R(t)>0,~\forall t\in[0,T]$ is a (potentially time-varying) symmetric matrix that penalizes excessive utilization of the control input. \medskip

\noindent\emph{\underline{Undesirable Populations:}} To penalize the population in an undesirable subspace $\ket\varphi$, it is possible to assign the incremental wavefunction cost
\begin{equation}
    \ell_\psi(\ket{\psi},t)=\tfrac12\mathfrak q(t)\bra{\psi} \overline\Gamma_{\varphi}\ket{\psi}
\end{equation}
where $\mathfrak q(t)$ is a is a (potentially time-varying) penalty weight and $\overline\Gamma_\varphi=\oprod{\varphi}{\varphi}$ is a positive definite matrix used to compute the projection of $\ket{\psi}$ onto $\ket\varphi$.
\medskip

\noindent\emph{\underline{No Penalties:}} If the wavefunction is allowed to evolve freely during the control period $t\in[0,T]$, the common choice for the incremental wavefunction cost is simply
\begin{equation}
    \ell_\psi(\ket{\psi},t)=0.
\end{equation}

\subsection{Quantum Gate Design}\label{ssec:gates}
The optimal control problem \eqref{eq:ocp_Quantum} can be used to design quantum gates by interpreting $\ket{\psi(t)}$ as a vector of orthonormal wavefunctions $\ket{\psi_i(t)}$ subject to the same control hamiltonian $\mathcal{H}_\psi(u)$. Given a target unitary $U$, let $\ket{U_i}$ denote its $i-$th column and let $\ket{e_i}$ be the $i$-th column of the identity matrix. Then, designing a quantum gate is equivalent to finding a $u(t)$ that simultaneously drives each element of the orthonormal basis from the initial condition $\ket{e_i}$ to the target $\ket{U_i}$. This can be achieved as follows.\medskip

\noindent\textbf{Schr\"odinger Equation}\\
Since each element of the orthonormal basis evolves under the same control Hamiltonian $H_\psi(u)$ starting from different initial conditions, \eqref{eq:Schrodinger} can be rewritten as
\begin{equation}
        \mathrm{i}\hbar\ket{\dot\psi_i(t)} = \mathcal H_\psi\bigl(u(t)\bigr)\ket{\psi_i(t)},\qquad \ket{\psi_i(0)}=\ket{e_i},
\end{equation}
for $i=1,\ldots,n$, where $n$ is the size of the unitary $U$. To do so, it is sufficient to define $\ket\psi$ as a column vector of all the $\ket{\psi_i}$ and $\mathcal{H}(u)=I_n\otimes\mathcal{H}_\psi(u)$, where $I_n$ is the identity matrix of size $n$ and $\otimes$ is the Kronecker product.\medskip

\noindent\textbf{Terminal Cost}\\
Since the relative phase between orthonormal basis vectors is crucial to the definition of a unitary, it is best to use the ``zero phase error'' terminal cost
\begin{equation}
    \mathcal J(\ket{\psi(T)})=\tfrac12\sum_{i=1}^n\left\|\ket{\psi_i(T)}-\ket{U_i}\right\|^2.
\end{equation}
Note that it is possible to design \emph{partial} unitary transforms by omitting terminal costs associated to basis vectors that are considered irrelevant. An example for this is given in Section \ref{ssec:GateExample}.\medskip

\noindent\textbf{Incremental Cost}\\
The incremental cost can be chosen as 
\begin{equation}
    \ell (\ket{\psi}\!,u,t\big)= \ell_u(u,t) + \sum_{i=1}^n\ell_i(\ket{\psi_i}\!,t),
\end{equation}
where each cost $\ell_i(\ket{\psi_i}\!,t)$ can be used to penalize the specific trajectory taken by each basis vector $\ket{\psi_i(t)}$.

\subsection{Equivalent Formulations}
Given the Schr\"odinger equation and a suitable cost function, the final step for implementing PRONTO is to reformulate the complex-valued optimal control problem \eqref{eq:ocp_Quantum} into the real-valued optimal control problem
\begin{subequations}\label{eq:ocp_original}
\begin{eqnarray}
\displaystyle \min \quad&&\displaystyle m(x(T))+\int_0^Tl(x(t),u(t),t)dt \label{eq:opt_cost}\\
\textrm{s.t.}\quad&& \dot x(t)=H(u(t))\,x(t),\qquad x(0)=x_0,\label{eq:opt_cstr}
\end{eqnarray}
\end{subequations}
where $x(t) \in \mathbb R^{2n}$ is the new state variable. As detailed in Appendix \ref{app:bijective}, this can be achieved without any loss of generality by applying the bijective mapping
\begin{equation}\label{eq:bijective}
    x=\begin{bmatrix}
    \mathrm{Re}(\ket{\psi})\\
    \mathrm{Im}(\ket{\psi})
    \end{bmatrix},\qquad \ket\psi=[\,I~~\mathrm{i}I\,]x,
\end{equation}
where $I$ is an identity matrix of size $n$. The real-valued optimal control problem \eqref{eq:ocp_original} is also equivalent to the Banach space optimization problem
\begin{equation}\label{eq:ocp_Banach}
    \min_{\xi\in\mathcal T} ~ h(\xi),
\end{equation}
where $\xi=[x(t),u(t)]$ is a state-and-input trajectory pair, 
\begin{equation}\label{eq:cost}
    h(\xi)=m(x(T))+\int_0^Tl(x(t),u(t),t)dt,
\end{equation}
is the cost functional, and
\begin{equation}
    \mathcal{T}=\{\xi~|~\dot x(t)=\mathcal H(u(t))\,x(t),~~x(0)=x_0\},\vspace{3pt}
\end{equation} 
is the trajectory manifold. The main interest in having the three equivalent formulations \eqref{eq:ocp_Quantum}, \eqref{eq:ocp_original}, and \eqref{eq:ocp_Banach} is that they each provide a slightly different insight onto the same problem.

\section{PRONTO for Quantum Systems}\label{sec:PRONTO}
As detailed in \cite{jay2022}, the idea behind Q-PRONTO 
is to turn the constrained optimization problem \eqref{eq:ocp_Banach} into an \emph{unconstrained} optimization problem
\begin{equation}\label{eq:simp_ocp}
    \min~ g(\eta),
\end{equation}
where $\eta=[\alpha(t),\mu(t)]$ is a pair of state-and-input curves that may not belong to the trajectory manifold $\mathcal T$. This is achieved by defining
\begin{equation}
    g(\eta)=h(\mathcal P(\eta)),
\end{equation}
where $\mathcal P(\eta)$ is an operator that projects any $\eta$ onto the manifold $\mathcal T$. This paper improves the convergence rate of \cite[Algorithm 1]{jay2022} by implementing a projection operator that relies not only on the input curve $\mu(t)$, but also on the state curve $\alpha(t)$. The updated method is detailed in the new algorithm provided in Appendix \ref{alg:Q-PRONTO}.\smallskip

The following subsections introduce the new projection operator and show how it can be used to compute the Newton descent direction. For the sake of brevity, algorithmic elements that remain unchanged with respect to \cite{jay2022}, i.e., the quasi-Newton step (lines 14--17) and the use of the Armijo step (lines 24--28) are not detailed in this paper.

\subsection{The Projection Operator}
Based on the original PRONTO formulation \cite{hauser2002projection,john2003}, given $\xi=[x(t),u(t)]$ and $\eta=[\alpha(t),\mu(t)]$, the projection $\xi=\mathcal{P}(\eta)$ is obtained by solving the differential equation
\begin{equation}\label{eq:simp_Proj}
    \left\{\begin{array}{l}
     \dot{x}(t)=H(u(t))\,x(t), \qquad\qquad\qquad~~x(0)= x_0,\\
         u(t) = \mu(t) - K_r(t)(x(t)-\alpha(t)),
    \end{array}\right. 
\end{equation}
where the regulator $K_r(t)$ is a time-varying feedback gain designed to ensure that trajectory $x(t)$ tracks the curve $\alpha(t)$ as closely as possible. By construction, this operator satisfies the following properties
\begin{itemize}
    \item Any pair of state-and-input curves $\eta$ is mapped onto the trajectory manifold, i.e., $\mathcal P(\eta)\in\mathcal T,~\forall \eta$;
    \item Any $\xi$ that is already in the trajectory manifold remains unaffected, i.e. $\mathcal P(\xi)=\xi,~\forall \xi\in\mathcal T$.
\end{itemize}
The design of $K_r(t)\neq0$ is addressed in Section \ref{sec:regulator}. For the case $K_r(t)=0$ the reader is referred to \cite{jay2022}.\smallskip

In view of implementing a Newton descent method, the remainder of this section will compute the first and second order approximations of the projection operator.\medskip

\noindent\textbf{First Order Approximation}\\
Given $\xi\in\mathcal T$, the first order approximation of $\mathcal P(\xi)$ along a general direction $\gamma$ yields $\zeta=D\mathcal{P}(\xi)\circ\gamma$, where the curve $\zeta=[z(t),v(t)]$ is obtained from $\gamma=[\beta(t),\nu(t)]$ by solving the differential equation
\begin{equation}\label{eq:FirstDer}
    \left\{\begin{array}{l}
     \dot{z}(t)=A_\xi(t)z(t)+B_\xi(t)v(t), \qquad\quad z(0)= 0,\\
         v(t) = \nu(t) - K_r(t)(z(t)-\beta(t)),
    \end{array}\right.
\end{equation}
with
\begin{subequations}\label{eq:LinDyn}
\begin{eqnarray}
    &&A_\xi(t)=H(u(t)),\\
    &&B_\xi(t)=\displaystyle\begin{bmatrix} \frac{\partial H}{\partial u_1}x(t)&\ldots& \frac{\partial H}{\partial u_m}x(t) \end{bmatrix},
\end{eqnarray}
\end{subequations}
representing to the first Fr\'echet derivative of $H(u)x$. With some abuse of notation, we refer to $[A_k(t),B_k(t)]$ as \eqref{eq:LinDyn} evaluated in $\xi_k=[x_k(t),u_k(t)]$. An interesting property of the projection operator is that \eqref{eq:FirstDer} acts as a projection of the curve $\gamma$ onto the tangent space of the trajectory manifold
\begin{equation}
    T_\xi\mathcal T=\{\zeta~|~\dot{z}(t)=A_\xi(t)z(t)+B_\xi(t)v(t),z(0)=0\}.
\end{equation}
Indeed, $\forall \xi\in\mathcal{T}$, it follows that
\begin{subequations}
\begin{eqnarray}
D\mathcal P(\xi)\circ\gamma&\in& T_\xi\mathcal T,\qquad\forall \gamma,\\
D\mathcal P(\xi)\circ\zeta&=&\zeta,\qquad\quad~\forall\zeta\in T_\xi\mathcal T.
\end{eqnarray}
\end{subequations}\smallskip

\noindent\textbf{Second Order Approximation}\\
Given $\xi\in\mathcal T$ and $\zeta\in T_\xi\mathcal T$, the second order approximation of $\mathcal P(\xi)$ yields $\omega=D^2\mathcal{P}(\xi)\circ(\zeta,\zeta)$, where $\omega=[y(t),w(t)]$ is obtained from the curve $\zeta=[z(t),v(t)]$ by solving the differential equation
\begin{equation}\label{eq:SecondDer}
    \left\{\begin{array}{l}
     \dot{y}(t)=A_\xi(t)y(t)+B_\xi(t)w(t)+\Lambda_{\xi}(t|\zeta)\\
         w(t) = - K_r(t)y(t),
    \end{array}\right.
\end{equation}
with $y(0)= 0$, where
\begin{equation*}
\begin{array}{rl}
    \Lambda_{\xi}(t|\zeta)=2&\!\!\displaystyle\sum_{i=1}^m \frac{\partial H}{\partial u_i}v_i(t)z(t)\\+&\!\!\displaystyle
    \sum_{i=1}^m\begin{bmatrix} \frac{\partial^2 H}{\partial u_1\partial u_i}x(t)&\ldots& \frac{\partial^2 H}{\partial u_m\partial u_i}x(t) \end{bmatrix}v_i(t)v(t),\\
\end{array}
\end{equation*}
is the first order approximation of $A_\xi(t)z(t)+B_\xi(t)v(t)$, linearized around $\xi$ and evaluated in the direction $\zeta$. 

\subsection{Newton Descent}
Given a solution estimate $\xi_k\in\mathcal T$, it is possible to compute a local (i.e. $\zeta\in T_\xi\mathcal T$) quadratic approximation of the cost function $g(\eta)$ by defining $\eta=\xi_k+\zeta$ and performing the second order expansion
\begin{equation}
    g(\eta)-g(\xi_k)\approx D g(\xi_k)\circ \zeta +\frac12D^2 g(\xi_k)\circ(\zeta,\zeta).
\end{equation}
As a result, the solution to \eqref{eq:simp_ocp} can be computed by performing the Newton step $\eta_k=\xi_k+\zeta_k$, where the local update $\zeta_k$ is obtained by solving the linear-quadratic optimal control problem
\begin{equation}\label{eq:simp_LQR}
    \min_{\zeta\in T_{\xi_k}\!\!\mathcal T} ~ D g(\xi_k)\circ \zeta +\frac12D^2 g(\xi_k)\circ(\zeta,\zeta).
\end{equation} 
The solution to \eqref{eq:ocp_Banach} can thus be computed using the projected Newton step
\begin{equation}
    \xi_{k+1}=\mathcal{P}(\xi_k+\zeta_k),
\end{equation}
The following subsections derive the first and second Fr\'echet derivative terms $Dg$ and $D^2g$.\medskip

\noindent \textbf{Linear Terms}\\
Since $g(\eta)=h(\mathcal P(\eta))$, it follows from the chain rule and the properties of the projection operator that
\begin{equation}\label{eq:first_der}
     D g(\xi_k)\circ \zeta=Dh(\mathcal{P}(\xi_k))\circ D\mathcal{P}(\xi_k)\circ \zeta=Dh(\xi_k)\circ \zeta.
\end{equation}
Evaluating the first Fr\'echet derivative of the cost \eqref{eq:cost} in $\xi_k=[x_k(t),u_k(t)]$ and composing it with $\zeta=[z(t),v(t)]$ leads to
\begin{equation}
    Dh(\xi_k)\circ\zeta=\pi_k z(T)+\!\!\int_0^Tq_k(t)z(t)+r_k(t)v(t)d\tau,
\end{equation}
with
\begin{subequations}\label{eq:LinCost}
\begin{eqnarray}
\pi_k&=&\nabla_x m(x_k(T)),\\
q_k(t)&=&\nabla_xl(x_k(t),u_k(t),t),\\
r_k(t)&=&\nabla_ul(x_k(t),u_k(t),t).
\end{eqnarray}
\end{subequations}

\noindent \textbf{Quadratic Terms}\\
Since $g(\eta)=h(\mathcal P(\eta))$, it follows from the chain rule and the properties of the projection operator that
\[
\begin{array}{rl}
     D^2 g(\xi_k)\!\circ\! (\zeta,\zeta)=&\!D^2h(\mathcal{P}(\xi_k))\!\circ\! (D\mathcal{P}(\xi_k)\!\circ\! \zeta,D\mathcal{P}(\xi_k)\!\circ\! \zeta)\\&
     \!\!\!\!+\,Dh(\mathcal{P}(\xi_k))\!\circ\! D^2\mathcal{P}(\xi_k)\!\circ\!(\zeta,\zeta)\\
     =&\!D^2h(\xi_k)\circ (\zeta,\zeta)+\,Dh(\xi_k)\circ\omega.
\end{array}
\]
Evaluating the second Fr\'echet derivative of \eqref{eq:cost} and composing it with $\zeta=[z(t),v(t)]$ leads to
\[
\begin{array}{rl}
     D^2h(\xi_k)\circ(\zeta,\zeta)=\! & z(T)^T\Pi_k z(T)+\ldots \\
     &\displaystyle \int_0^T\!\begin{bmatrix}z(t)\\ v(t)\end{bmatrix}^{\!T}\!\!\begin{bmatrix}\bar Q_k(t) & \bar S_k(t)\\ \bar S_k^T(t)&\bar R_k(t)\end{bmatrix}\!\!\begin{bmatrix}z(t)\\ v(t)\end{bmatrix}
    dt,
\end{array}
\]
with 
\begin{subequations}\label{eq:quasi_Newton}
\begin{eqnarray}
\Pi_k&=&\nabla^2_{xx} m(x_k(T)),\\
\bar Q_k(t)&=&\nabla^2_{xx}l(x_k(t),u_k(t),t),\\
\bar S_k(t)&=&\nabla^2_{xu}l(x_k(t),u_k(t),t),\\
\bar R_k(t)&=&\nabla^2_{uu}l(x_k(t),u_k(t),t).
\end{eqnarray}
\end{subequations}
As for the other term, the composition of $Dh(\xi_k)$ with $\omega=[y(t),w(t)]$ yields
\begin{equation}\label{eq:SecondOrder}
    Dh(\xi_k)\circ\omega=\pi_k y(T)+\!\!\int_0^Tq_k(t)y(t)+r_k(t)w(t)d\tau.    
\end{equation}
Unfortunately, this equation does not feature an explicit dependency on the optimization variable $\zeta$.  
To overcome this challenge, consider the state transition matrix
$\Phi(t,\tau)$ satisfying
\begin{equation}
    \dot\Phi(t,\tau)=(A(t)-B(t)K_r(t))\Phi(t,\tau).
\end{equation}
Then, the solution to \eqref{eq:SecondDer} is
\begin{equation}
    y(t)=\int_0^t\Phi(t,\tau)\Lambda_{\xi}(\tau|\zeta)ds.
\end{equation}
\begin{figure*}
\begin{equation}\label{eq:LQR}
\begin{array}{rl}
    \min ~&\displaystyle \pi_kz(T)+\frac12z(T)\Pi_kz(T)+ \displaystyle\int_0^T\begin{bmatrix}q_k(t)~~r_k(t)\end{bmatrix}\begin{bmatrix}z(t)\\v(t)\end{bmatrix}+\frac12\begin{bmatrix}z(t)\\ v(t)\end{bmatrix}^{\!T}\!\begin{bmatrix}Q_k(t) & S_k(t)\\S_k^T(t)&R_k(t)\end{bmatrix}\begin{bmatrix}z(t)\\v(t)\end{bmatrix}
    dt\\~\\
\textrm{s.t.}~& \dot z(t)=A_k(t)z(t)+B_k(t)v(t),\qquad\qquad z(0)=0,
\end{array}
\end{equation}
\end{figure*}
Since $w(t)=-K_r(t)y(t)$, equation \eqref{eq:SecondOrder} becomes
\[
\begin{array}{l}
    \displaystyle Dh(\xi_k)\circ \omega=\pi_k y(T)+\int_0^T\!\!\!\bigl(q_k(t)-r_k(t)K_r(t)\bigr)\,y(t)dt\\\displaystyle=
    \pi_k y(T)\!+\!\int_0^T\!\!\!\!\int_0^t\!\!\!\bigl(q_k(t)\!-\!r_k(t)K_r(t)\bigr)\Phi(t,\tau)\Lambda_{\xi}(\tau|\zeta)d\tau\,dt\\
    \displaystyle=\pi_k y(T)\!+\!\int_0^T\!\!\!\!\int_s^T\!\!\! \bigl(q_k(t)\!-\!r_k(t)K_r(t)\bigr)\Phi(t,\tau)\Lambda_{\xi}(\tau|\zeta)dt\,d\tau\\
    \displaystyle=\int_0^T\chi_k^T(\tau)\Lambda_{\xi}(\tau|\zeta)d\tau,
\end{array}
\]
where
\[
    \chi_k^T(\tau)=\pi_k\Phi(T,\tau) +\int_\tau^T (q_k(t)-r_k(t)K_r(t))\Phi(t,\tau) dt
\]
is an adjoint variable that evolves backwards in time.
Due to the properties of state transition matrices, $\chi_k(t)$ can now be obtained by solving the differential equation
\begin{equation}\label{eq:adjoint}
    -\dot\chi_k= (A_k-B_k K_r)^T\chi_k+(q_k-r_kK_r)^T\!\!\!,
\end{equation}
with final conditions $\chi(T)=\pi_k$.
\begin{rmk}
Given $K_r(t)=0$, the adjoint dynamics \eqref{eq:adjoint} reduce to the co-state dynamics featured in the Krotov method \cite[Eq. (30)]{KrotovToolkit2019}, i.e., $-\dot\chi_k= A_k^T\chi_k+q_k^T$. As such, the adjoint variable $\chi_k(t)$ can be interpreted as a co-state that has been stabilized by the regulator $K_r(t)$.
\end{rmk}

The direct dependency from $\zeta=[z(t),v(t)]$ is then shown to be
\[
    \chi_k^T(t)\Lambda_\xi(t|\zeta)=\!\begin{bmatrix}z(t)\\v(t)\end{bmatrix}^{\!T}\!\!\begin{bmatrix}0 & \tilde S_k(t)\\\tilde S_k^T(t)&\tilde R_k(t)\end{bmatrix}\!\!\begin{bmatrix}z(t)\\v(t)\end{bmatrix},
\]
with
\[
\begin{array}{rl}
    \tilde S_k(t)&=\displaystyle\begin{bmatrix} \frac{\partial H}{\partial u_i}\chi_k(t)&\ldots& \frac{\partial H}{\partial u_m}\chi_k(t) \end{bmatrix},\\~\\
    \tilde R_k(t)&=\displaystyle\begin{bmatrix}\chi_k^T(t) H_{11}(t)x_k(t)&\ldots&\chi_k^T(t) H_{m1}(t)x_k(t)\\
    \vdots&\ddots&\vdots\\ \chi_k^T(t) H_{1m}(t)x_k(t)&\ldots&\chi_k^T(t) H_{mm}(t)x_k(t) \end{bmatrix},
\end{array}
\]
and $H_{ij}(t)=\frac{\partial^2 H}{\partial u_i\partial u_j}$.
As a result, the evaluation of \eqref{eq:simp_LQR} leads to the Linear-Quadratic Optimal Control Problem (LQ-OCP) in equation \eqref{eq:LQR}, with
\begin{subequations}\label{eq:Newton}
\begin{eqnarray}
 Q_k(t)&=&\bar Q_k(t),\\
S_k(t)&=&\bar S_k(t)+\tilde S_k(t),\\
R_k(t)&=&\bar R_k(t)+\tilde R_k(t).
\end{eqnarray}
\end{subequations}

The main interest in the LQ-OCP \eqref{eq:LQR} is that its solution can be computed explicitly \cite{Opt_ctrl}. To do so, we first solve the Differential Riccati Equation
\begin{equation}\label{eq:DRE}
    \left\{\begin{array}{rll}
    -\dot P =&A^T_kP\!+\!PA_k-K_o^TR_kK_o+Q_k,\\
    -\dot p=&(A_k-B_kK_o)^Tp-K_o^Tr_k^T+q_k^T,\\
    K_o =& R^{-1}_k(B_k^TP+S_k^T),\\
    v_o = & R_k^{-1}(B^T_kp+r_k^T),
    \end{array}\right.
\end{equation}
with final conditions $P(T)\!=\!\Pi_k$ and $p(T)=\pi_k$.
Then, the update $\zeta_k$ can be obtained by solving
\begin{equation}\label{eq:update}
    \left\{\begin{array}{rll}
    \dot z_k(t) =&A_k(t)z_k(t)+B_k(t) v_k(t),~~ & z(0)=0,\\
    v_k(t)=&-v_o(t)-K_o(t)z_k(t).
    \end{array}\right.
\end{equation}
Additional algorithmic considerations, such as using the Armijo rule to limit the Newton step size, or replacing $(Q_k,S_k,R_k)$ with $(\bar{Q}_k,\bar{S}_k,\bar{R}_k)$ whenever \eqref{eq:LQR} does not admit a solution, are detailed in prior work \cite{jay2022}.

\section{Regulator Design}\label{sec:regulator}
Since the projection operator \eqref{eq:simp_Proj} is defined up to a time-varying feedback gain $K_r(t)$, the objective of this section is to design a regulator such that the trajectory $\xi=\mathcal{P}(\eta)$ tracks $\eta$ ``as closely as possible''. To provide a more formal metric for success, consider the error coordinates $\Delta x(t) = x(t) - \alpha(t)$ and $\Delta u(t) = u(t) - \mu(t)$. Then, given arbitrary cost matrices $Q_r(t)\geq0$, $R_r(t)>0$, and $\Pi_r\geq0$, the optimal feedback gain $K_r(t)$ can be defined in relationship to the LQ-OCP
\begin{subequations}\label{eq:LQOCP_reg}
\begin{eqnarray}
\displaystyle \min~~ && \tfrac{1}{2}\Delta x(T)^T\Pi_r\Delta x(T) +\ldots \nonumber \\&& \tfrac{1}{2}\!\int^T_0 \!\!\begin{bmatrix}\Delta x(t)\\ \Delta u(t)\end{bmatrix}^{\!T}\!\!\begin{bmatrix}Q_r(t)&0\\ 0& R_r(t)\end{bmatrix} \!\begin{bmatrix}\Delta x(t)\\ \Delta u(t)\end{bmatrix}\!dt  \\ 
\nonumber \\
\textrm{s.t.}~~&& \Delta\dot{x} = A_\eta(t)\Delta x(t) + B_\eta(t)\Delta u(t),
\end{eqnarray}
\end{subequations}
where $$A_\eta(t)=H(\mu(t)),~~B_\eta(t)=\displaystyle\begin{bmatrix} \frac{\partial H}{\partial \mu_1}(t)\alpha(t)&\ldots& \frac{\partial H}{\partial \mu_m}\alpha(t) \end{bmatrix}\!,$$
are the linearization of the systems dynamics around $\eta(t)=[\alpha(t),\mu(t)]$. Indeed, as detailed in \cite{Opt_ctrl}, the optimal controller for the LQ-OCP \eqref{eq:LQOCP_reg} is $\Delta u(t)=-K_r(t)\Delta x(t)$, where the feedback gain $K_r(t)$ is obtained by solving the Differential Riccati Equation
\begin{equation}\label{eq:DRE_reg}
    \left\{\begin{array}{rll}
    -\dot P_r =&A^T_\eta P_r+P_rA_\eta-K_r^TR_rK_r+Q_r,\\
    K_r =& R^{-1}_rB^T_\eta P_r,
    \end{array}\right.
\end{equation}
with final conditions $P_r(T)=\Pi_r$. Thus, the design of $K_r(t)$ follows directly from the choice of the cost matrices in the LQ-OCP \eqref{eq:LQOCP_reg}. \medskip

\noindent\emph{\underline{Global Phase Projection:}} In line of principle, any positive definite cost function would lead to a suitable regulator $K_r(t)$. In this regard, the simplest choice is
\begin{equation}\label{eq:cost_global}
    Q_r(t)=I_{2n},\qquad R_r(t)=c_R I_m, \qquad \Pi_r= c_P I_{2n},
\end{equation}
with $c_R>0$ and $c_P\geq0$, which leads to the cost function
\begin{equation}
    \frac{c_P}2\|\Delta x(T)\|^2+\frac12\int_0^T\|\Delta x(t)\|^2+c_R\|\Delta u(t)\|^2dt.
\end{equation}
Under this metric, the feedback gain $K_r(t)$ minimizes the weighted error between the trajectory $\xi=[x(t),u(t)]$ and $\eta=[\alpha(t),\mu(t)]$. This metric, however, may be overly restrictive for quantum systems since the global phase between two wave functions is often irrelevant.\medskip

\noindent\emph{\underline{Arbitrary Phase Projection:}} To design a projection operator that is phase-agnostic, consider the wavefunctions 
\begin{equation}\label{eq:bijective2}
    \ket\psi=[\,I~~\mathrm{i}I\,]x,\qquad\ket\phi=[\,I~~\mathrm{i}I\,]\alpha.
\end{equation}
These two wavefunctions are equal up to an arbitrary phase if $\bra\psi(I-\oprod\phi\phi)\ket\psi=0$. As detailed in Appendix \ref{app:bijective}, this condition is equivalent to $x^T\Phi(\alpha) x=0$, with

\begin{equation}
    \Phi(\alpha)=\begin{bmatrix}
    I-\alpha_r\alpha_r^T-\alpha_i\alpha_i^T & 0\\
    0 & I-\alpha_r\alpha_r^T-\alpha_i\alpha_i^T\end{bmatrix},
\end{equation}
where $\alpha_r=[I~~0]\,\alpha$ and $\alpha_i=[0~~I]\,\alpha$. To avoid penalizing global phase errors between $x(t)$ and $\alpha(t)$, it is therefore sufficient to assign
\begin{subequations}\label{eq:cost_arbitrary}
\begin{eqnarray}
 Q_r(t)&=&\Phi(\alpha(t)),\\
 R_r(t)&=&c_R I_m,\\
\Pi_r&=&c_P\Phi(\alpha(T)),
\end{eqnarray}
\end{subequations}
with $c_R>0$ and $c_P\geq0$.

\section{Numerical Examples}\label{sec:NumEx}
\subsection{Beam Splitter}
\begin{figure}
    \centering
    \includegraphics[width=0.5\textwidth]{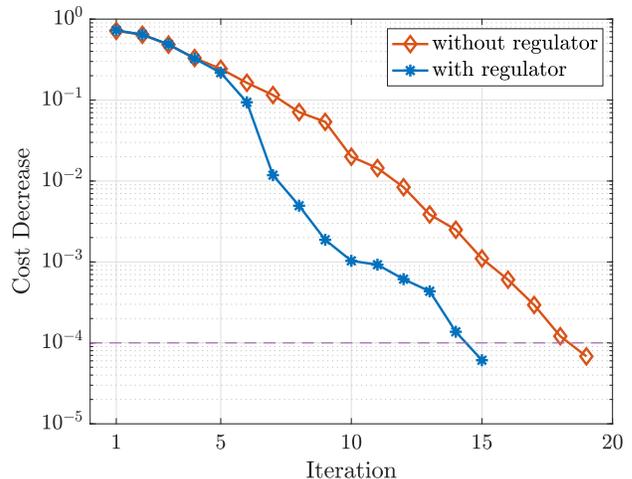}
    \caption{Value of the cost decrease $-Dg(\eta_k)\cdot\zeta_k$ at each iteration. The purple dashed line is the exit condition $\texttt{tol} = 10^{-4}$. The introduction of a regulator gain $K_r(t)\neq0$ causes the solver to converge faster ($20\%$ less iterations).}
    \label{fig_descnet}
\end{figure}

We now apply Q-PRONTO to a atom cloud beam splitter, which is the first stage of an atom-based interferometer \cite{shao2023acc}. Consider the Schr\"{o}dinger equation
\begin{equation}
    \mathrm{i} \hbar \ket{\dot{\psi}} = (H_0 + \sin{u(t)}H_1 + (1-\cos{u(t)})H_2)\ket{\psi},
\end{equation}
where the matrices  $H_i$ are derived in \cite{CSM2023}. Define $\ket{n}$ as the $n^{th}$ eigenvector of $H_0$, we want to perform a state transition from $\ket{1}$ to $\ket{4}$. To achieve this goal, we minimize the cost function
\begin{equation}\label{eq:cost_split}
    \frac{1}{2} \bra{\psi(T)} P \ket{\psi(T)} + \int^T_0 \frac{r}{2} \|u(t)\|^2 dt,
\end{equation}
where $P = I - \ketbra{4}{4}$, $r = 0.01$ is a penalty on the energy of control input $u(t)$, and $T = 10~ \omega_R^{-1}$ is the time horizon. This optimal control problem can then be rewritten as
\begin{subequations}
\begin{eqnarray}
\displaystyle \min~~ && \frac{1}{2}x(T)^T P x(T) + \int^T_0 \frac{r}{2} \|u(t)\|^2 dt\\
\textrm{s.t.}~~&& \dot{x} = (H_0 + \sin{u}H_1 + (1-\cos{u})H_2).
\end{eqnarray}
\end{subequations}
To compare Q-PRONTO with the regulator \eqref{eq:DRE_reg} and without regulator (i.e., $K_r(t) = 0 ~\forall t \in [0,T]$), we initialize our solver using the same guess 
\begin{equation}
    u_0(t) = 0.5\sin{t}.
\end{equation}
Given the exit condition $\texttt{tol} = 10^{-4}$, Figure \ref{fig_descnet} illustrates the value of $\texttt{tol}$ at each iteration. The method with regulator meets the desired \texttt{tol} after 15 iterations while the method without regulator uses 19 iterations.  

\begin{figure}
    \centering
    \includegraphics[width=0.5\textwidth]{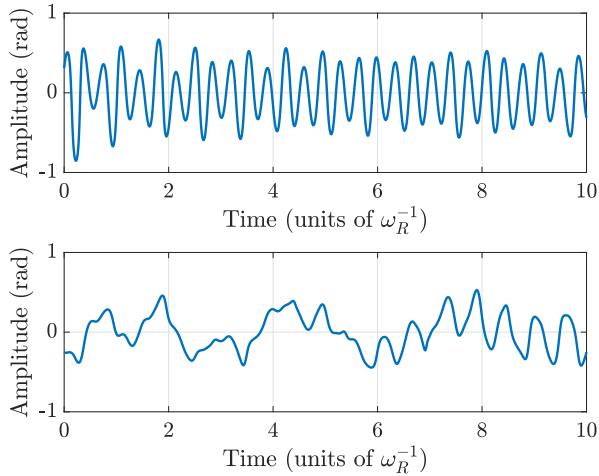}
    \caption{Optimal control inputs obtained without regulator (\textbf{Top}: $K_r(t)=0$) and with regulator (\textbf{Bottom}: $K_r(t)\neq0$). The latter clearly features a lower frequency content, making it preferable from the implementation perspective.}
    \label{fig_input}
\end{figure}

Figure \ref{fig_input} shows the optimal control laws $u(t)$ obtained from both methods. We can see that the two optimal control laws are different from each other, meaning that the two solvers have converged to a different local minimum of the quantum optimal control problem. 

\begin{figure}
    \centering
    \includegraphics[width=0.5\textwidth]{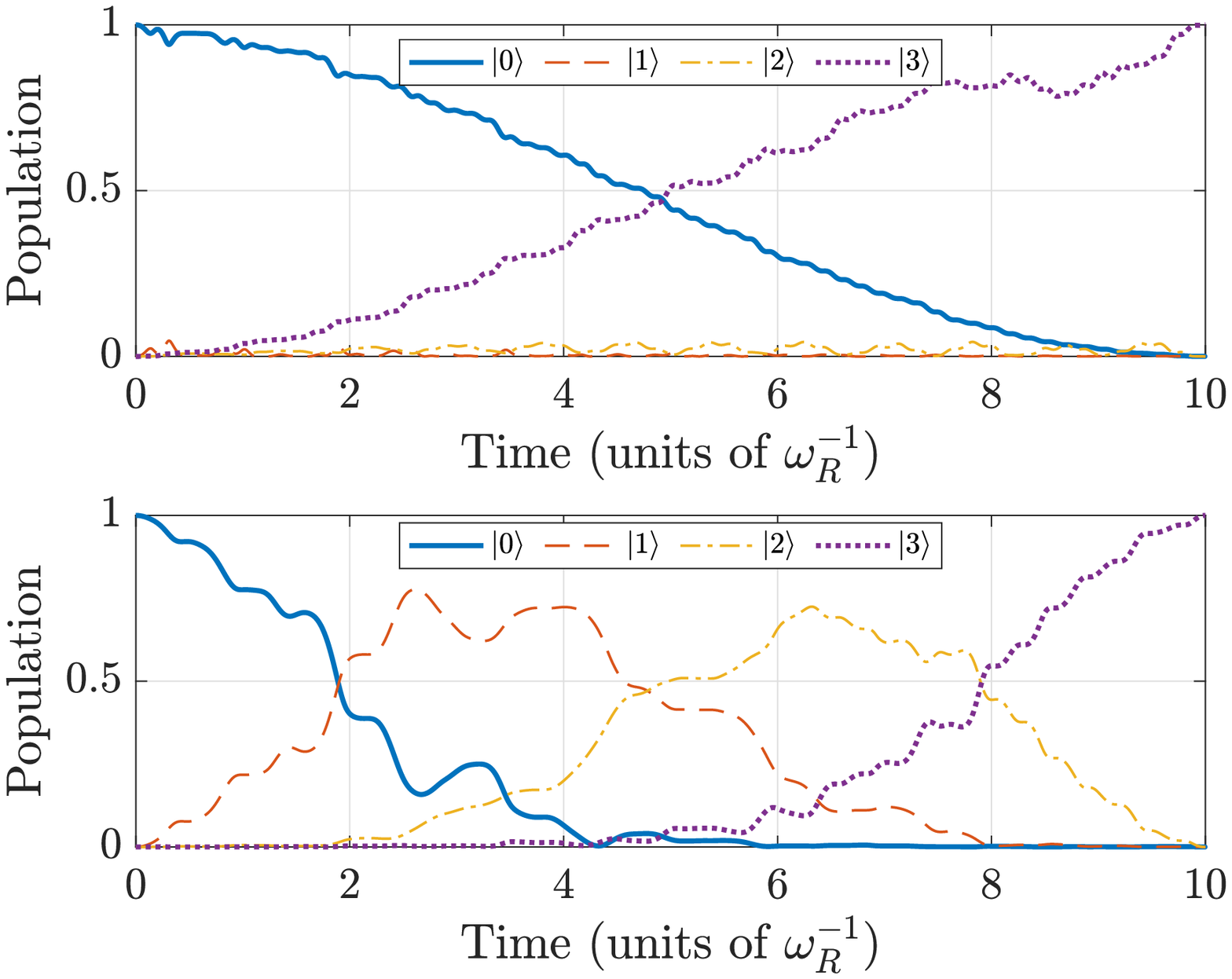}
    \caption{Population dynamics obtained without regulator (\textbf{Top}: $K_r(t)=0$) and with regulator (\textbf{Bottom}: $K_r(t)\neq0$). The former forces the direct transition $\ket0\!\to\!\ket3$, whereas the latter achieves the natural progression $\ket0\!\to\!\ket1\!\to\!\ket2\!\to\!\ket3$.}
    \label{fig_all_state}
\end{figure}

Figure \ref{fig_all_state} shows how the population of each eigenstate 
\begin{equation}
    F_n(t) = \abs{\langle\psi(t)|n\rangle}^2,\qquad n = 0,1,2,3,
\end{equation}
evolves over time. The two optimal control laws have substantially different behaviors. The optimal control law obtained with the regulator sequentially steers the system from the initial state $\ket{0}$, to the second eigenstate $\ket{1}$, then the third eigenstate $\ket{2}$, and finally the target state $\ket{3}$. However, the optimal control law obtained without the regulator tries to steer the system from the initial state $\ket{0}$ to the target state $\ket{3}$ directly. This explains why the two optimal control laws are so different. The one obtained with the regulator starts with a lower frequency (since the lower eigenstate represents lower energy) and gradually increases in frequency as it reaches a higher eigenstates. Meanwhile, the one obtained without the regulator maintains a high frequency throughout the whole time horizon because it forces the system to transition directly into the highest eigenstate $\ket{3}$. \smallskip

\begin{figure}
    \centering
    \includegraphics[width=0.5\textwidth]{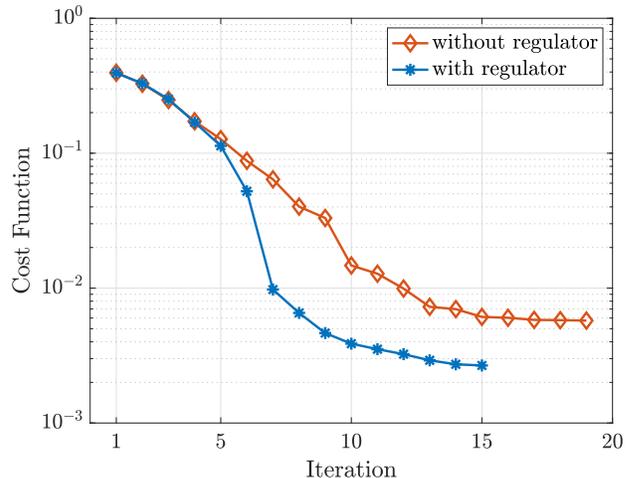}
    \caption{Value of the cost function \eqref{eq:cost_split} at each iteration of the two solvers. The solution obtained with the regulator has a lower cost than the one obtained without a regulator. The introduction of a regulator gain $K_r(t)\neq0$ causes the solver to converge to a better ($50\%$ cost reduction) local minimum.}
    \label{fig_cost}
\end{figure}

To identify which control law is preferable, Figure \ref{fig_cost} illustrates the value of the cost function \eqref{eq:cost_split} at each iteration. Unsurprisingly, the solution obtained with the regulator achieve a lower cost value ($2.7\times 10^{-3}$) compared to the solution obtained without the regulator ($5.7\times 10^{-3}$). Physically, this can be explained by the fact that the former attains the target state-by-state in a more ``natural'' way (i.e., transitioning between energy levels), whereas the latter takes a direct path that requires more energy.\smallskip

This numerical example highlights an unexpected benefit of adding a non-zero feedback term into the projection operator: selecting $K_r(t)\neq0$ not only improves the convergence rate, but can also steer the towards a better minimizer of the quantum optimal control problem.

\subsection{State-to-State Transfer in a Lambda System}
We now consider a 3-level system in a $\Lambda$ configuration, whose eigenstates are $\ket{0}$, $\ket{1}$ and $\ket{2}$ with energy levels $E_0<E_2<E_1$. We wish to find the optimal control inputs that steer the system from $\ket{0}$ to $\ket{1}$ without exciting $\ket{2}$. As detailed in \cite{KrotovToolkit2019}, the Schr\"{o}dinger equation for this system is 

\begin{equation}\label{eq:real_lambda}
\begin{array}{cc}
     \mathrm{i}\hbar \ket{\dot{\psi}} = H(u)\ket{\psi},\\
    H(u) = H_0\! +\! u_1 H_{P1} \!+\! u_2 H_{P2}\! +\! u_3 H_{S1} \!+\! u_4 H_{S2},
\end{array}
\end{equation}
with
\begin{equation}
\begin{array}{rl}
    H_0&=\Delta_P\oprod{0}{0} + \Delta_S\oprod{1}{1}   \\
    H_{P1}&=-1/2(\oprod{0}{2} + \oprod{2}{0}), \\
    H_{P2}&=-\mathrm{i}/2(\oprod{0}{2} - \oprod{2}{0}), \\
    H_{S1}&=-1/2(\oprod{2}{1} + \oprod{1}{2}), \\
    H_{S2}&=-\mathrm{i}/2(\oprod{2}{1} - \oprod{1}{2}).
\end{array}
\end{equation}
Note that, for this problem, we care about the global phase error between the target state $\ket{1}$ and the final state $\ket{\psi(T)}$, so we pick a phase-sensitive terminal cost
\begin{equation}
    \mathcal J(\ket{\psi(T)})=1-\mathrm{Re}\left(\iprod{\psi(T)}{1}\right).
\end{equation}
For the same reason, we pick the global phase projection regulator \eqref{eq:cost_global} to compute $K_r(t)$, with $C_R\!=\!1$ and $C_P\!=\!1$. Thanks to the regulator, 
our initial guess does not even need to be a trajectory. To show this, we initialize the Q-PRONTO solver 
using a smooth hyperbolic tangent function that connects $\ket{\psi_0}=\ket{0}$ to $\ket{\psi_T}=\ket{1}$, meaning
$$\ket{\psi_0(t)}=\ket{\psi_T-\psi_0}(\tanh(\tfrac{2\pi}{T}t-\pi)+1)/2+\ket{\psi_0}.$$ 
Although this curve is not a feasible trajectory for the system, the projection operator turns it into a suitable initial guess belonging on the trajectory manifold. For the sake of comparison, we propose two different incremental costs. For both cases, we set the time horizon $T = 5$ and the exit condition $\texttt{tol} = 10^{-6}$.\medskip 

\begin{figure}
    \centering
    \includegraphics[width=0.5\textwidth]{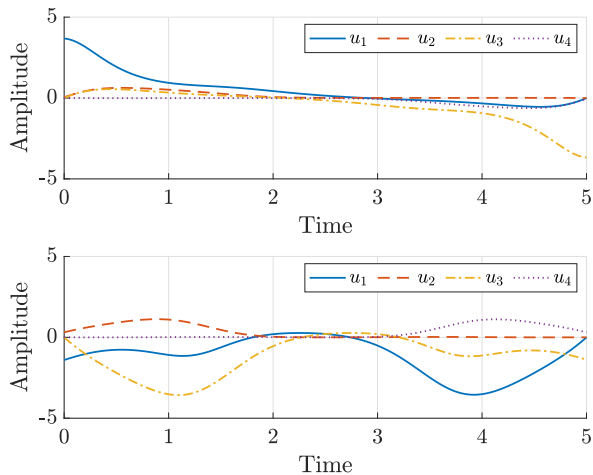}
    \caption{Optimal control input sequences obtained when the objectives are to minimize the control effort (\textbf{Top}) or prevent the system from populating $\ket 2$ (\textbf{Bottom}).}
    \label{fig_u_lambda}
\end{figure}
\begin{figure}
    \centering
    \includegraphics[width=0.5\textwidth]{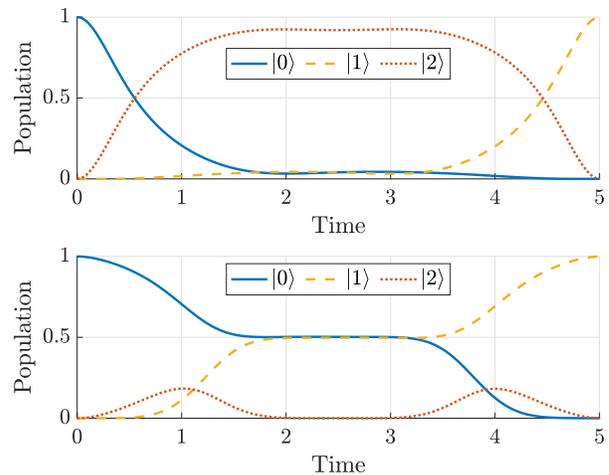}
    \caption{Population dynamics obtained under the optimal control law that minimizes the control effort (\textbf{Top}) or prevents the system from populating $\ket 2$ (\textbf{Bottom}).}
    \label{fig_p_lambda}
\end{figure}

\noindent\textbf{Minimize Control Effort}\\
First, we consider a incremental cost that only penalizes the control effort, i.e. 
\begin{equation}\label{eq:lambda_cost_1}
    \ell (\ket{\psi}\!,u,t\big)=\tfrac12u^T \mathcal R(t)\, u,
\end{equation}
where $$\mathcal R(t)=0.1\begin{bmatrix}0.01 & 0 & 0 & 0\\ 0 & r(t)+1.1 & 0 & 0\\ 0 & 0 & 0.01 & 0\\ 0 & 0 & 0 & -r(t)+1.1
\end{bmatrix},$$
and $r(t)=\tanh(2t-T)$. Figure \ref{fig_u_lambda}a shows the optimal control input $u(t)$ and Figure \ref{fig_p_lambda}a shows the population evolves in time. Although almost all population goes to the target state at the end ($t=5$), we can clearly observe that there is over $90\%$ of the population going through the state $\ket{2}$ for $t \in (1,4)$. This behavior may be undesirable for some applications. The solver took $11$ iterations to obtain the solution.\medskip 

\noindent\textbf{Minimize Undesirable Population}\\
To avoid populating the $\ket{2}$ state as much as possible, consider the incremental cost
\begin{equation}\label{eq:lambda_cost_2}
    \ell (\ket{\psi}\!,u,t\big) = \tfrac12u^T \mathcal R(t)\, u + \tfrac12\mathfrak q\bra{\psi} \overline\Gamma_{2}\ket{\psi},    
\end{equation}
where $\mathcal R(t)$ is the same as \eqref{eq:lambda_cost_1} and $\mathfrak q = 0.1$, $\overline\Gamma_{2} = \oprod{2}{2}$ define a new cost that penalizes $\ket{2}$.\smallskip

Figures \ref{fig_u_lambda}b and \ref{fig_p_lambda}b  show the optimal control input $u(t)$ and the population $\ket{\psi(t)}$, respectively. In this case, the system successfully transitions from $\ket0$ to $\ket 1$ while maintaining a limited ($<20\%$) presence in the undesirable population $\ket2$. The solver took $19$ iterations to obtain the solution.

\subsection{X-Gate for a Fluxonium Qubit}\label{ssec:GateExample}
We now show how to design a Pauli-X gate for a fluxonium qubit using Q-PRONTO. The Hamiltonian of the fluxonium qubit \cite{manucharyan2009fluxonium} is given by
\begin{equation}
    H_0 = -4E_C\partial^2_{\phi}-E_J\cos{(\phi-\varphi_{\text{ext}}})+\frac{1}{2}E_L\phi^2,
\end{equation}
where $E_C$ is the charging energy, $E_J$ is the Josephson energy, $E_L$ is the inductive energy, and $\varphi_{\text{ext}}$ is the external flux in dimensionless form.
Consider a truncated 3-level fluxonium qubit, whose Hamiltonian can be written as
\begin{equation}\label{eq:qubit}
    \mathrm{i}\hbar \ket{\dot{\psi}} = \left(\!\begin{bmatrix}
        E_0 & 0 & 0\\
        0 & E_1 & 0\\
        0 & 0 & E_2
    \end{bmatrix} \!+\! u \begin{bmatrix}
        0 & \Omega_{01} & \Omega_{02}\\
        \Omega_{01} & 0 & \Omega_{12}\\
        \Omega_{02} & \Omega_{12} & 0
    \end{bmatrix}\!\right)\!\! \ket{\psi},
\end{equation}
where $E_0 = 0$, $E_1 = 1~\text{GHz}$, $E_2 = 5~\text{GHz}$ is the corresponding energy levels for the eigenstates $\ket{0}$, $\ket{1}$ and $\ket{2}$; $\Omega_{01} = 0.1$, $\Omega_{12} = 0.5$, $\Omega_{02} = 0.3$ is the effort to transfer between eigenstates. Since $\ket{0}$ and $\ket{1}$ are the logical states for this qubit, we wish to design a Pauli X-gate: an optimal control input $u(t)$ so that it steers $\ket{0}$ to $\ket{1}$, while simultaneously steering $\ket{1}$ to $\ket{0}$. The optimal control problem can then be written as
\begin{subequations}
\begin{eqnarray}
\displaystyle \min~~ && \|\psi_0(T)\!-\!\ket{1}\|^2 + \|\psi_1(T)\!-\!\ket{0}\|^2 + \int^T_0 \!\!\!l(\ket{\psi},u,t) dt\nonumber\\
\textrm{s.t.}~~&& \ket{\dot{\psi_0}} = H(u)\ket{\psi_0}, \qquad\ket{\psi_0(0)}=\ket{0},\\
~~&& \ket{\dot{\psi_1}} = H(u)\ket{\psi_1}, \qquad\ket{\psi_1(0)}=\ket{1}.
\end{eqnarray}
\end{subequations}
We propose two different incremental costs, but for both cases, we set the time horizon $T=10$ ns and use the same initial guess pulse
$$u_0(t)=\frac{\pi}{T} e^{\frac{-(t-T/2)^2}{T^2}}\cos{2\pi t}.$$
The exit condition is $\texttt{tol} = 10^{-4}$.
\begin{figure}
    \centering
    \includegraphics[width=0.5\textwidth]{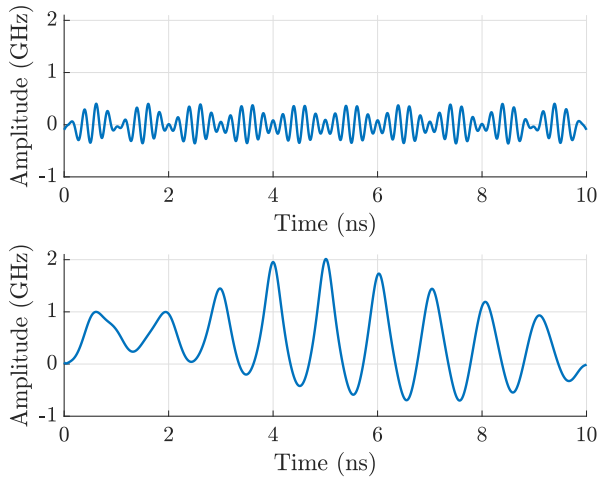}
    \caption{Optimal control input sequences obtained when the objectives are to minimize the control effort (\textbf{Top}) or to prevent the system from populating $\ket 2$ (\textbf{Bottom}).}
    \label{fig_u_xgate}
\end{figure}

\begin{figure}
    \centering
    \includegraphics[width=0.5\textwidth]{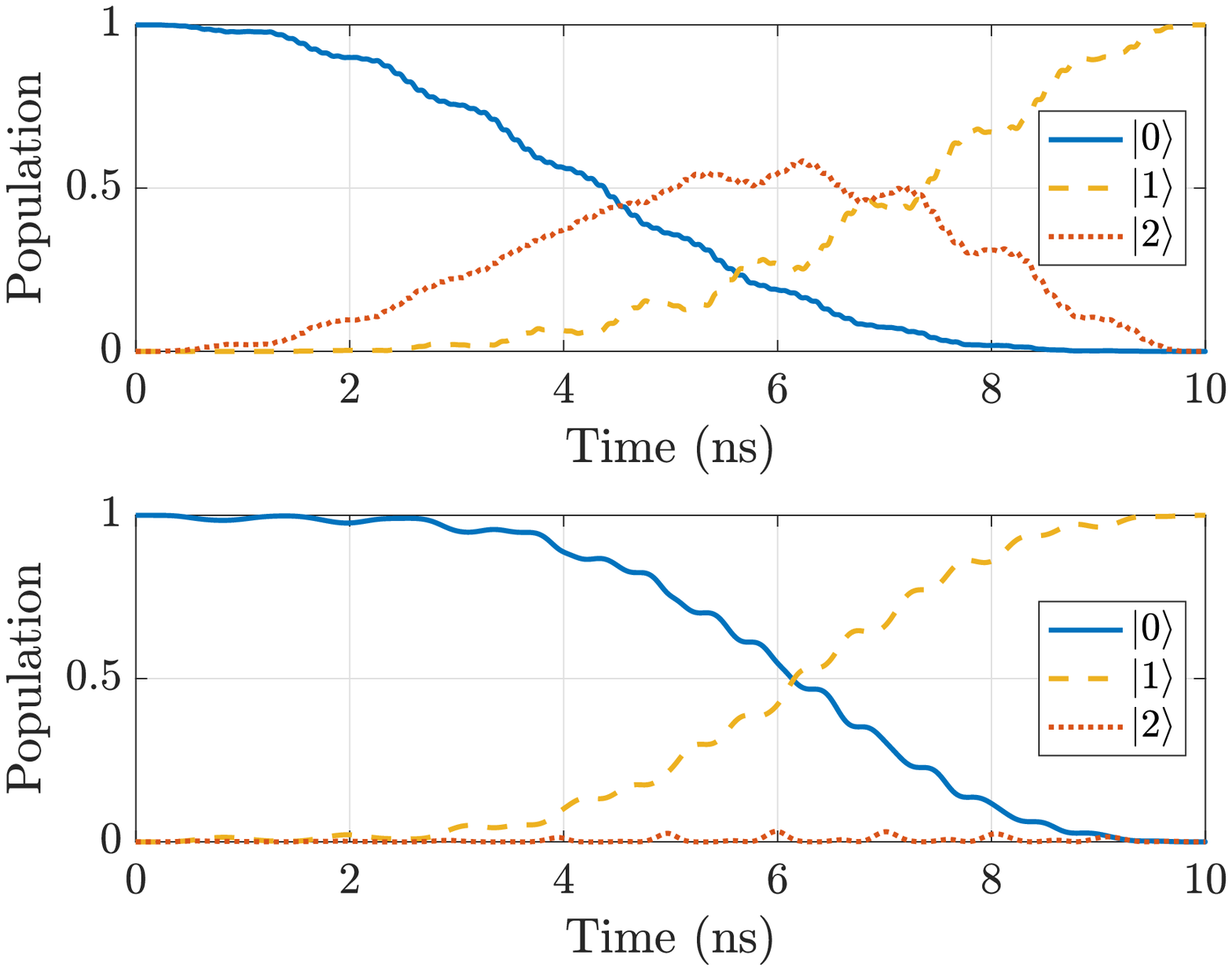}
    \caption{Population dynamics of the $\ket0\!\to\!\ket1$ transition under the optimal control law that minimizes the control effort (\textbf{Top}) or prevents the system from populating $\ket 2$ (\textbf{Bottom}). The transient peak of $\ket2$ is reduced from $.58$ to $.03$.}
    \label{fig_p_xgate_1}
\end{figure}

\begin{figure}
    \centering
    \includegraphics[width=0.5\textwidth]{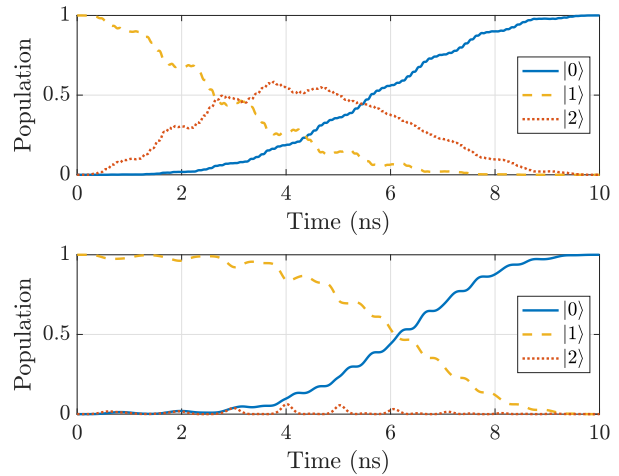}
    \caption{Population dynamics of the $\ket1\!\to\!\ket0$ transition under the optimal control law that minimizes the control effort (\textbf{Top}) or prevents the system from populating $\ket 2$ (\textbf{Bottom}). The behavior roughly mirrors the $\ket0\!\to\!\ket1$ transition.}
    \label{fig_p_xgate_2}
\end{figure}

\noindent\textbf{Minimize Control Effort}\\
First, we consider a incremental cost that only penalizes the control effort, i.e. 
\begin{equation}\label{eq:qubit_cost_1}
    \ell (\ket{\psi}\!,u,t\big)=\tfrac12u^T u.
\end{equation}
Figure \ref{fig_u_xgate}a shows the optimal control input $u(t)$ and Figures \ref{fig_p_xgate_1}a--\ref{fig_p_xgate_2}a  show how the two basis vectors $\ket{\psi_0}$ and $\ket{\psi_1}$ evolve over time. Although the quantum gate fidelity \cite{gate_fidelity} 
$$\mathcal{F}=(d + |\Tr(UU_{\text{targ}}^{\dagger})|^2)/(d^2+d)$$
is greater than $99.9\%$, we can clearly observe that over $50\%$ of the population transitions through the state $\ket{2}$ during the process. The solver took $10$ iterations to obtain the solution.  \medskip

\noindent\textbf{Minimize Undesirable Population}\\
To avoid populating the $\ket{2}$ state as much as possible, consider the incremental cost
\begin{equation}\label{eq:qubit_cost_2}
    \ell (\ket{\psi}\!,u,t\big) = \tfrac12u^Tu + \tfrac12\mathfrak q(\bra{\psi_0} \overline\Gamma_{2}\ket{\psi_0}+\bra{\psi_1} \overline\Gamma_{2}\ket{\psi_1}),    
\end{equation}
where $\mathfrak q = 0.3$ and $\overline\Gamma_{2} = \oprod{2}{2}$ penalize $\ket{2}$.\smallskip

Figures \ref{fig_u_xgate}b and \ref{fig_p_xgate_1}b--\ref{fig_p_xgate_2}b  show the optimal control input $u(t)$ and the basis vector dynamics $\ket{\psi_0(t)}$, $\ket{\psi_1(t)}$, respectively. In this case, the gate fidelity remains greater than $99.9\%$ while maintaining a limited ($<3\%$) presence in the $\ket{2}$ population during the whole process. However, the control amplitude becomes much higher, with the maximum amplitude reaching $2 \text{GHz}$. Note that the tradeoff between maximum input amplitude and maximum $\ket2$ population can be explored by tuning $\mathfrak q\geq0$. The solver took $4$ iterations to obtain the solution.

\section{Conclusions}
This paper augments the Quantum Projection Operator-based Newton method for Trajectory Optimization (Q-PRONTO) with a quantum-specific regulator designed to stabilize the system trajectories to the spherical manifold. The method is not only faster due to the presence of the regulator, but may also converge to better local minima compared to the unregulated case. 

Additionally, the paper introduces new capabilities to Q-PRONTO, such as the ability to incorporate time-varying cost functions and to avoid undesirable populations during the transient. Future research directions include the design of unitary gates, control of open quantum systems, and in-depth comparisons with existing optimization methods. Q-PRONTO is set to be released as an open source Julia package over the next year.\smallskip

\noindent {\em Acknowledgements}
The authors would like to thank Joshua Combes, Andr\'as Gyenis, Murray Holland, and Liang-Ying Chih for their valued collaboration and helpful insights. This work is supported by the NSF QII–TAQS award number 1936303.



\begin{thebibliography}{19}%
\makeatletter
\providecommand \@ifxundefined [1]{%
 \@ifx{#1\undefined}
}%
\providecommand \@ifnum [1]{%
 \ifnum #1\expandafter \@firstoftwo
 \else \expandafter \@secondoftwo
 \fi
}%
\providecommand \@ifx [1]{%
 \ifx #1\expandafter \@firstoftwo
 \else \expandafter \@secondoftwo
 \fi
}%
\providecommand \natexlab [1]{#1}%
\providecommand \enquote  [1]{``#1''}%
\providecommand \bibnamefont  [1]{#1}%
\providecommand \bibfnamefont [1]{#1}%
\providecommand \citenamefont [1]{#1}%
\providecommand \href@noop [0]{\@secondoftwo}%
\providecommand \href [0]{\begingroup \@sanitize@url \@href}%
\providecommand \@href[1]{\@@startlink{#1}\@@href}%
\providecommand \@@href[1]{\endgroup#1\@@endlink}%
\providecommand \@sanitize@url [0]{\catcode `\\12\catcode `\$12\catcode
  `\&12\catcode `\#12\catcode `\^12\catcode `\_12\catcode `\%12\relax}%
\providecommand \@@startlink[1]{}%
\providecommand \@@endlink[0]{}%
\providecommand \url  [0]{\begingroup\@sanitize@url \@url }%
\providecommand \@url [1]{\endgroup\@href {#1}{\urlprefix }}%
\providecommand \urlprefix  [0]{URL }%
\providecommand \Eprint [0]{\href }%
\providecommand \doibase [0]{https://doi.org/}%
\providecommand \selectlanguage [0]{\@gobble}%
\providecommand \bibinfo  [0]{\@secondoftwo}%
\providecommand \bibfield  [0]{\@secondoftwo}%
\providecommand \translation [1]{[#1]}%
\providecommand \BibitemOpen [0]{}%
\providecommand \bibitemStop [0]{}%
\providecommand \bibitemNoStop [0]{.\EOS\space}%
\providecommand \EOS [0]{\spacefactor3000\relax}%
\providecommand \BibitemShut  [1]{\csname bibitem#1\endcsname}%
\let\auto@bib@innerbib\@empty
\bibitem [{\citenamefont {Brif}\ \emph {et~al.}(2010)\citenamefont {Brif},
  \citenamefont {Chakrabarti},\ and\ \citenamefont {Rabitz}}]{brif2010control}%
  \BibitemOpen
  \bibfield  {author} {\bibinfo {author} {\bibfnamefont {C.}~\bibnamefont
  {Brif}}, \bibinfo {author} {\bibfnamefont {R.}~\bibnamefont {Chakrabarti}},\
  and\ \bibinfo {author} {\bibfnamefont {H.}~\bibnamefont {Rabitz}},\
  }\bibfield  {title} {\bibinfo {title} {Control of quantum phenomena: past,
  present and future},\ }\href@noop {} {\bibfield  {journal} {\bibinfo
  {journal} {New Journal of Physics}\ }\textbf {\bibinfo {volume} {12}},\
  \bibinfo {pages} {075008} (\bibinfo {year} {2010})}\BibitemShut {NoStop}%
\bibitem [{\citenamefont {Dong}\ and\ \citenamefont
  {Petersen}(2010)}]{dong2010quantum}%
  \BibitemOpen
  \bibfield  {author} {\bibinfo {author} {\bibfnamefont {D.}~\bibnamefont
  {Dong}}\ and\ \bibinfo {author} {\bibfnamefont {I.~R.}\ \bibnamefont
  {Petersen}},\ }\bibfield  {title} {\bibinfo {title} {Quantum control theory
  and applications: a survey},\ }\href@noop {} {\bibfield  {journal} {\bibinfo
  {journal} {IET control theory \& applications}\ }\textbf {\bibinfo {volume}
  {4}},\ \bibinfo {pages} {2651} (\bibinfo {year} {2010})}\BibitemShut
  {NoStop}%
\bibitem [{\citenamefont {Altafini}\ and\ \citenamefont
  {Ticozzi}(2012)}]{altafini2012modeling}%
  \BibitemOpen
  \bibfield  {author} {\bibinfo {author} {\bibfnamefont {C.}~\bibnamefont
  {Altafini}}\ and\ \bibinfo {author} {\bibfnamefont {F.}~\bibnamefont
  {Ticozzi}},\ }\bibfield  {title} {\bibinfo {title} {Modeling and control of
  quantum systems: An introduction},\ }\href@noop {} {\bibfield  {journal}
  {\bibinfo  {journal} {IEEE Transactions on Automatic Control}\ }\textbf
  {\bibinfo {volume} {57}},\ \bibinfo {pages} {1898} (\bibinfo {year}
  {2012})}\BibitemShut {NoStop}%
\bibitem [{\citenamefont {Glaser}\ \emph {et~al.}(2015)\citenamefont {Glaser},
  \citenamefont {Boscain}, \citenamefont {Calarco}, \citenamefont {Koch},
  \citenamefont {K{\"o}ckenberger}, \citenamefont {Kosloff}, \citenamefont
  {Kuprov}, \citenamefont {Luy}, \citenamefont {Schirmer}, \citenamefont
  {Schulte-Herbr{\"u}ggen} \emph {et~al.}}]{glaser2015training}%
  \BibitemOpen
  \bibfield  {author} {\bibinfo {author} {\bibfnamefont {S.~J.}\ \bibnamefont
  {Glaser}}, \bibinfo {author} {\bibfnamefont {U.}~\bibnamefont {Boscain}},
  \bibinfo {author} {\bibfnamefont {T.}~\bibnamefont {Calarco}}, \bibinfo
  {author} {\bibfnamefont {C.~P.}\ \bibnamefont {Koch}}, \bibinfo {author}
  {\bibfnamefont {W.}~\bibnamefont {K{\"o}ckenberger}}, \bibinfo {author}
  {\bibfnamefont {R.}~\bibnamefont {Kosloff}}, \bibinfo {author} {\bibfnamefont
  {I.}~\bibnamefont {Kuprov}}, \bibinfo {author} {\bibfnamefont
  {B.}~\bibnamefont {Luy}}, \bibinfo {author} {\bibfnamefont {S.}~\bibnamefont
  {Schirmer}}, \bibinfo {author} {\bibfnamefont {T.}~\bibnamefont
  {Schulte-Herbr{\"u}ggen}}, \emph {et~al.},\ }\bibfield  {title} {\bibinfo
  {title} {Training schr{\"o}dinger’s cat: quantum optimal control},\
  }\href@noop {} {\bibfield  {journal} {\bibinfo  {journal} {The European
  Physical Journal D}\ }\textbf {\bibinfo {volume} {69}},\ \bibinfo {pages} {1}
  (\bibinfo {year} {2015})}\BibitemShut {NoStop}%
\bibitem [{\citenamefont {Khaneja}\ \emph {et~al.}(2001)\citenamefont
  {Khaneja}, \citenamefont {Brockett},\ and\ \citenamefont
  {Glaser}}]{khaneja2001time}%
  \BibitemOpen
  \bibfield  {author} {\bibinfo {author} {\bibfnamefont {N.}~\bibnamefont
  {Khaneja}}, \bibinfo {author} {\bibfnamefont {R.}~\bibnamefont {Brockett}},\
  and\ \bibinfo {author} {\bibfnamefont {S.~J.}\ \bibnamefont {Glaser}},\
  }\bibfield  {title} {\bibinfo {title} {Time optimal control in spin
  systems},\ }\href@noop {} {\bibfield  {journal} {\bibinfo  {journal}
  {Physical Review A}\ }\textbf {\bibinfo {volume} {63}},\ \bibinfo {pages}
  {032308} (\bibinfo {year} {2001})}\BibitemShut {NoStop}%
\bibitem [{\citenamefont {Khaneja}\ \emph {et~al.}(2005)\citenamefont
  {Khaneja}, \citenamefont {Reiss}, \citenamefont {Kehlet}, \citenamefont
  {Schulte-Herbr{\"u}ggen},\ and\ \citenamefont {Glaser}}]{khaneja2005optimal}%
  \BibitemOpen
  \bibfield  {author} {\bibinfo {author} {\bibfnamefont {N.}~\bibnamefont
  {Khaneja}}, \bibinfo {author} {\bibfnamefont {T.}~\bibnamefont {Reiss}},
  \bibinfo {author} {\bibfnamefont {C.}~\bibnamefont {Kehlet}}, \bibinfo
  {author} {\bibfnamefont {T.}~\bibnamefont {Schulte-Herbr{\"u}ggen}},\ and\
  \bibinfo {author} {\bibfnamefont {S.~J.}\ \bibnamefont {Glaser}},\ }\bibfield
   {title} {\bibinfo {title} {Optimal control of coupled spin dynamics: design
  of nmr pulse sequences by gradient ascent algorithms},\ }\href@noop {}
  {\bibfield  {journal} {\bibinfo  {journal} {Journal of magnetic resonance}\
  }\textbf {\bibinfo {volume} {172}},\ \bibinfo {pages} {296} (\bibinfo {year}
  {2005})}\BibitemShut {NoStop}%
\bibitem [{\citenamefont {Sklarz}\ and\ \citenamefont
  {Tannor}(2002)}]{sklarz2002loading}%
  \BibitemOpen
  \bibfield  {author} {\bibinfo {author} {\bibfnamefont {S.~E.}\ \bibnamefont
  {Sklarz}}\ and\ \bibinfo {author} {\bibfnamefont {D.~J.}\ \bibnamefont
  {Tannor}},\ }\bibfield  {title} {\bibinfo {title} {Loading a bose-einstein
  condensate onto an optical lattice: An application of optimal control theory
  to the nonlinear schr{\"o}dinger equation},\ }\href@noop {} {\bibfield
  {journal} {\bibinfo  {journal} {Physical Review A}\ }\textbf {\bibinfo
  {volume} {66}},\ \bibinfo {pages} {053619} (\bibinfo {year}
  {2002})}\BibitemShut {NoStop}%
\bibitem [{\citenamefont {Reich}\ \emph {et~al.}(2012)\citenamefont {Reich},
  \citenamefont {Ndong},\ and\ \citenamefont {Koch}}]{reich2012monotonically}%
  \BibitemOpen
  \bibfield  {author} {\bibinfo {author} {\bibfnamefont {D.~M.}\ \bibnamefont
  {Reich}}, \bibinfo {author} {\bibfnamefont {M.}~\bibnamefont {Ndong}},\ and\
  \bibinfo {author} {\bibfnamefont {C.~P.}\ \bibnamefont {Koch}},\ }\bibfield
  {title} {\bibinfo {title} {Monotonically convergent optimization in quantum
  control using krotov's method},\ }\href@noop {} {\bibfield  {journal}
  {\bibinfo  {journal} {The Journal of chemical physics}\ }\textbf {\bibinfo
  {volume} {136}},\ \bibinfo {pages} {104103} (\bibinfo {year}
  {2012})}\BibitemShut {NoStop}%
\bibitem [{\citenamefont {de~Fouquieres}\ \emph {et~al.}(2011)\citenamefont
  {de~Fouquieres}, \citenamefont {Schirmer}, \citenamefont {Glaser},\ and\
  \citenamefont {Kuprov}}]{de2011second}%
  \BibitemOpen
  \bibfield  {author} {\bibinfo {author} {\bibfnamefont {P.}~\bibnamefont
  {de~Fouquieres}}, \bibinfo {author} {\bibfnamefont {S.~G.}\ \bibnamefont
  {Schirmer}}, \bibinfo {author} {\bibfnamefont {S.~J.}\ \bibnamefont
  {Glaser}},\ and\ \bibinfo {author} {\bibfnamefont {I.}~\bibnamefont
  {Kuprov}},\ }\bibfield  {title} {\bibinfo {title} {Second order gradient
  ascent pulse engineering},\ }\href@noop {} {\bibfield  {journal} {\bibinfo
  {journal} {Journal of Magnetic Resonance}\ }\textbf {\bibinfo {volume}
  {212}},\ \bibinfo {pages} {412} (\bibinfo {year} {2011})}\BibitemShut
  {NoStop}%
\bibitem [{\citenamefont {Eitan}\ \emph {et~al.}(2011)\citenamefont {Eitan},
  \citenamefont {Mundt},\ and\ \citenamefont {Tannor}}]{eitan2011optimal}%
  \BibitemOpen
  \bibfield  {author} {\bibinfo {author} {\bibfnamefont {R.}~\bibnamefont
  {Eitan}}, \bibinfo {author} {\bibfnamefont {M.}~\bibnamefont {Mundt}},\ and\
  \bibinfo {author} {\bibfnamefont {D.~J.}\ \bibnamefont {Tannor}},\ }\bibfield
   {title} {\bibinfo {title} {Optimal control with accelerated convergence:
  Combining the krotov and quasi-newton methods},\ }\href@noop {} {\bibfield
  {journal} {\bibinfo  {journal} {Physical Review A}\ }\textbf {\bibinfo
  {volume} {83}},\ \bibinfo {pages} {053426} (\bibinfo {year}
  {2011})}\BibitemShut {NoStop}%
\bibitem [{\citenamefont {Shao}\ \emph {et~al.}(2022)\citenamefont {Shao},
  \citenamefont {Combes}, \citenamefont {Hauser},\ and\ \citenamefont
  {Nicotra}}]{jay2022}%
  \BibitemOpen
  \bibfield  {author} {\bibinfo {author} {\bibfnamefont {J.}~\bibnamefont
  {Shao}}, \bibinfo {author} {\bibfnamefont {J.}~\bibnamefont {Combes}},
  \bibinfo {author} {\bibfnamefont {J.}~\bibnamefont {Hauser}},\ and\ \bibinfo
  {author} {\bibfnamefont {M.~M.}\ \bibnamefont {Nicotra}},\ }\bibfield
  {title} {\bibinfo {title} {Projection-operator-based newton method for the
  trajectory optimization of closed quantum systems},\ }\href@noop {}
  {\bibfield  {journal} {\bibinfo  {journal} {Physical Review A}\ }\textbf
  {\bibinfo {volume} {105}},\ \bibinfo {pages} {032605} (\bibinfo {year}
  {2022})}\BibitemShut {NoStop}%
\bibitem [{\citenamefont {Hauser}(2002)}]{hauser2002projection}%
  \BibitemOpen
  \bibfield  {author} {\bibinfo {author} {\bibfnamefont {J.}~\bibnamefont
  {Hauser}},\ }\bibfield  {title} {\bibinfo {title} {A projection operator
  approach to the optimization of trajectory functionals},\ }\href@noop {}
  {\bibfield  {journal} {\bibinfo  {journal} {IFAC Proceedings Volumes}\
  }\textbf {\bibinfo {volume} {35}},\ \bibinfo {pages} {377} (\bibinfo {year}
  {2002})}\BibitemShut {NoStop}%
\bibitem [{\citenamefont {Hauser}(2003)}]{john2003}%
  \BibitemOpen
  \bibfield  {author} {\bibinfo {author} {\bibfnamefont {J.}~\bibnamefont
  {Hauser}},\ }\bibfield  {title} {\bibinfo {title} {On the computation of
  optimal state transfers with application to the control of quantum spin
  systems},\ }in\ \href {https://doi.org/10.1109/ACC.2003.1243395} {\emph
  {\bibinfo {booktitle} {Proceedings of the 2003 American Control Conference,
  2003.}}},\ Vol.~\bibinfo {volume} {3}\ (\bibinfo {year} {2003})\ pp.\
  \bibinfo {pages} {2169--2174 vol.3}\BibitemShut {NoStop}%
\bibitem [{\citenamefont {Goerz}\ \emph {et~al.}(2019)\citenamefont {Goerz},
  \citenamefont {Basilewitsch}, \citenamefont {Gago-Encinas}, \citenamefont
  {Krauss}, \citenamefont {Horn}, \citenamefont {Reich},\ and\ \citenamefont
  {Koch}}]{KrotovToolkit2019}%
  \BibitemOpen
  \bibfield  {author} {\bibinfo {author} {\bibfnamefont {M.~H.}\ \bibnamefont
  {Goerz}}, \bibinfo {author} {\bibfnamefont {D.}~\bibnamefont {Basilewitsch}},
  \bibinfo {author} {\bibfnamefont {F.}~\bibnamefont {Gago-Encinas}}, \bibinfo
  {author} {\bibfnamefont {M.~G.}\ \bibnamefont {Krauss}}, \bibinfo {author}
  {\bibfnamefont {K.~P.}\ \bibnamefont {Horn}}, \bibinfo {author}
  {\bibfnamefont {D.~M.}\ \bibnamefont {Reich}},\ and\ \bibinfo {author}
  {\bibfnamefont {C.~P.}\ \bibnamefont {Koch}},\ }\bibfield  {title} {\bibinfo
  {title} {Krotov: A {Python} implementation of {Krotov's} method for quantum
  optimal control},\ }\href {https://doi.org/10.21468/SciPostPhys.7.6.080}
  {\bibfield  {journal} {\bibinfo  {journal} {SciPost Phys.}\ }\textbf
  {\bibinfo {volume} {7}},\ \bibinfo {pages} {80} (\bibinfo {year}
  {2019})}\BibitemShut {NoStop}%
\bibitem [{\citenamefont {Anderson}\ and\ \citenamefont
  {Moore}(2007)}]{Opt_ctrl}%
  \BibitemOpen
  \bibfield  {author} {\bibinfo {author} {\bibfnamefont {B.~D.~O.}\
  \bibnamefont {Anderson}}\ and\ \bibinfo {author} {\bibfnamefont {J.~B.}\
  \bibnamefont {Moore}},\ }\bibfield  {title} {\bibinfo {title} {{Optimal
  Control: Linear Quadratic Methods}},\ }\href@noop {} {\bibfield  {journal}
  {\bibinfo  {journal} {Dover Books on Engineering}\ } (\bibinfo {year}
  {2007})}\BibitemShut {NoStop}%
\bibitem [{\citenamefont {Shao}\ \emph {et~al.}(2023)\citenamefont {Shao},
  \citenamefont {Chih}, \citenamefont {Naris}, \citenamefont {Holland},\ and\
  \citenamefont {Nicotra}}]{shao2023acc}%
  \BibitemOpen
  \bibfield  {author} {\bibinfo {author} {\bibfnamefont {J.}~\bibnamefont
  {Shao}}, \bibinfo {author} {\bibfnamefont {L.-Y.}\ \bibnamefont {Chih}},
  \bibinfo {author} {\bibfnamefont {M.}~\bibnamefont {Naris}}, \bibinfo
  {author} {\bibfnamefont {M.}~\bibnamefont {Holland}},\ and\ \bibinfo {author}
  {\bibfnamefont {M.~M.}\ \bibnamefont {Nicotra}},\ }\bibfield  {title}
  {\bibinfo {title} {Application of quantum optimal control to shaken lattice
  interferometry},\ }in\ \href
  {https://doi.org/10.23919/ACC55779.2023.10156455} {\emph {\bibinfo
  {booktitle} {2023 American Control Conference (ACC)}}}\ (\bibinfo {year}
  {2023})\ pp.\ \bibinfo {pages} {4593--4598}\BibitemShut {NoStop}%
\bibitem [{\citenamefont {Nicotra}\ \emph {et~al.}(2023)\citenamefont
  {Nicotra}, \citenamefont {Shao}, \citenamefont {Combes}, \citenamefont
  {Theurkauf}, \citenamefont {Axelrad}, \citenamefont {Chih}, \citenamefont
  {Holland}, \citenamefont {Zozulya}, \citenamefont {LeDesma}, \citenamefont
  {Mehling} \emph {et~al.}}]{CSM2023}%
  \BibitemOpen
  \bibfield  {author} {\bibinfo {author} {\bibfnamefont {M.~M.}\ \bibnamefont
  {Nicotra}}, \bibinfo {author} {\bibfnamefont {J.}~\bibnamefont {Shao}},
  \bibinfo {author} {\bibfnamefont {J.}~\bibnamefont {Combes}}, \bibinfo
  {author} {\bibfnamefont {A.~C.}\ \bibnamefont {Theurkauf}}, \bibinfo {author}
  {\bibfnamefont {P.}~\bibnamefont {Axelrad}}, \bibinfo {author} {\bibfnamefont
  {L.-Y.}\ \bibnamefont {Chih}}, \bibinfo {author} {\bibfnamefont
  {M.}~\bibnamefont {Holland}}, \bibinfo {author} {\bibfnamefont {A.~A.}\
  \bibnamefont {Zozulya}}, \bibinfo {author} {\bibfnamefont {C.~K.}\
  \bibnamefont {LeDesma}}, \bibinfo {author} {\bibfnamefont {K.}~\bibnamefont
  {Mehling}}, \emph {et~al.},\ }\bibfield  {title} {\bibinfo {title} {Modeling
  and control of ultracold atoms trapped in an optical lattice: An
  example-driven tutorial on quantum control},\ }\href@noop {} {\bibfield
  {journal} {\bibinfo  {journal} {IEEE Control Systems Magazine}\ }\textbf
  {\bibinfo {volume} {43}},\ \bibinfo {pages} {28} (\bibinfo {year}
  {2023})}\BibitemShut {NoStop}%
\bibitem [{\citenamefont {Manucharyan}\ \emph {et~al.}(2009)\citenamefont
  {Manucharyan}, \citenamefont {Koch}, \citenamefont {Glazman},\ and\
  \citenamefont {Devoret}}]{manucharyan2009fluxonium}%
  \BibitemOpen
  \bibfield  {author} {\bibinfo {author} {\bibfnamefont {V.~E.}\ \bibnamefont
  {Manucharyan}}, \bibinfo {author} {\bibfnamefont {J.}~\bibnamefont {Koch}},
  \bibinfo {author} {\bibfnamefont {L.~I.}\ \bibnamefont {Glazman}},\ and\
  \bibinfo {author} {\bibfnamefont {M.~H.}\ \bibnamefont {Devoret}},\
  }\bibfield  {title} {\bibinfo {title} {Fluxonium: Single cooper-pair circuit
  free of charge offsets},\ }\href@noop {} {\bibfield  {journal} {\bibinfo
  {journal} {Science}\ }\textbf {\bibinfo {volume} {326}},\ \bibinfo {pages}
  {113} (\bibinfo {year} {2009})}\BibitemShut {NoStop}%
\bibitem [{\citenamefont {Pedersen}\ \emph {et~al.}(2007)\citenamefont
  {Pedersen}, \citenamefont {M{\o}ller},\ and\ \citenamefont
  {M{\o}lmer}}]{gate_fidelity}%
  \BibitemOpen
  \bibfield  {author} {\bibinfo {author} {\bibfnamefont {L.~H.}\ \bibnamefont
  {Pedersen}}, \bibinfo {author} {\bibfnamefont {N.~M.}\ \bibnamefont
  {M{\o}ller}},\ and\ \bibinfo {author} {\bibfnamefont {K.}~\bibnamefont
  {M{\o}lmer}},\ }\bibfield  {title} {\bibinfo {title} {Fidelity of quantum
  operations},\ }\href@noop {} {\bibfield  {journal} {\bibinfo  {journal}
  {Physics Letters A}\ }\textbf {\bibinfo {volume} {367}},\ \bibinfo {pages}
  {47} (\bibinfo {year} {2007})}\BibitemShut {NoStop}%
\end{thebibliography}%
%

\appendix

\section{Wavefunction to State Mapping}\label{app:bijective}
For the reader's convenience, this section provides some background on how to use the bijective mapping \eqref{eq:bijective} to seamlessly transition between the \eqref{eq:ocp_Quantum} and \eqref{eq:ocp_original}. For the purpose of this section, let $a\in\mathbb R^n,~b\in\mathbb R^n$ be two real column vectors and define
\begin{equation}\label{eq:bijective3} 
    x=\begin{bmatrix}a\\b
    \end{bmatrix},\qquad \ket\psi=a+\mathrm{i}b
\end{equation}
so that \eqref{eq:bijective} is satisfied. Then, given the real matrices $A\in\mathbb R^{n\times n},~B\in\mathbb R^{n\times n}$ we note that
\begin{equation}
 \ket{\psi'}=(A+\mathrm{i}B)\ket\psi=(Aa-Bb)+\mathrm{i}(Ab+Ba)
\end{equation}
satisfies the bijective mapping \eqref{eq:bijective} for
\begin{equation}
x'=\begin{bmatrix}A & -B\\ B & ~~A
    \end{bmatrix}x=\begin{bmatrix}Aa-Bb\\Ab+Ba
    \end{bmatrix}.
\end{equation}
As such, the induced matrix mapping
\begin{equation}\label{eq:inducedmap}
    M=\begin{bmatrix}
    \mathrm{Re}(\mathcal M) & -\mathrm{Im}(\mathcal M)\\
    \mathrm{Im}(\mathcal M) & ~~ \mathrm{Re}(\mathcal M)
    \end{bmatrix},\qquad \mathcal M=[\,I~~\mathrm{i}I\,]M
\end{equation}
ensures that the state vector $x'=Mx$ remains consistent with the wavefunction $\ket{\psi'}=\mathcal{M}\ket\psi$. Based on these considerations, we show the following.

\subsection{Schr\"odinger Equation}
The Schr\"odinger equation $\ket{\dot\psi}=\mathcal H\ket\psi$ can be rewritten as $\dot x = H x$ with
\begin{equation}
    H=\begin{bmatrix}
    ~~0 & I~\\
    -I & 0~
    \end{bmatrix}\begin{bmatrix}
    \mathrm{Re}(\mathcal H) & -\mathrm{Im}(\mathcal H)\\
    \mathrm{Im}(\mathcal H) & ~~ \mathrm{Re}(\mathcal H)
    \end{bmatrix}.
\end{equation}
This can be shown by applying the induced matrix mapping \eqref{eq:inducedmap} to the equation $\ket{\dot\psi}=-\mathrm{i}\mathcal H\ket\psi$. Note that, if $\mathcal{H}$ is Hermitian (i.e., $\mathcal H^\dagger=\mathcal H$), $\mathrm{i}\mathcal H$ is skew-Hermitian, which makes $H$ skew-symmetric (i.e., $H^T=-H$). This is sufficient to ensure that $\|x(t)\|$ remains constant, since
\begin{equation}
    \frac d{dt}\|x\|^2=\dot x^Tx+x^T\dot x=x^T(H^T+H)x=0.
\end{equation}

\subsection{Quadratic Costs}
Although it is possible to find an equivalent mapping for any cost function, this paper only addresses quadratic cost functions
\begin{equation}
    c = \bra\psi \mathcal Q \ket\psi,
\end{equation}
where $\mathcal Q=A+\mathrm{i}B$ is poisitive semi-definite and Hermitian (i.e., $A$ symmetric and $B$ skew-symmetric). Using \eqref{eq:bijective3}, it is possible to show that 
\begin{equation}
    c = (a^T-\mathrm{i}b^T)(A+\mathrm{i}B)(a+\mathrm{i}b)=a^T\!Aa+b^T\!Ab.
\end{equation}
Since the same output can also be obtained by computing 
\begin{equation}
    c=x^T\begin{bmatrix}A & 0\\ 0 & A\end{bmatrix}x,
\end{equation}
it is possible to define the induced quadratic cost matrix
\begin{equation}
    Q=\begin{bmatrix}\mathrm{Re}(\mathcal Q) & 0\\ 0 & \mathrm{Re}(\mathcal Q)\end{bmatrix}.
\end{equation}
For example, given $\mathcal Q=\oprod\psi\psi$, we obtain
\[
    \mathcal Q=(a+\mathrm{i}b)(a^T-\mathrm{i}b^T)=(aa^T+bb^T)+\mathrm{i}(ba^T-ab^T),
\]
which entails
\begin{equation}
    Q=\begin{bmatrix}aa^T+bb^T & 0\\ 0 & aa^T+bb^T\end{bmatrix}.
\end{equation}

\section{Q-PRONTO Algorithm}\label{alg:Q-PRONTO}
\begin{algorithm}[H]
\caption{~~Q-PRONTO}
\begin{algorithmic}[1]
    \vspace{5 pt}\Statex \begin{center}
      \textbf{Input}\end{center} 
      \Statex \emph{Initial Guess}
      \State $\eta_0=(\alpha_0(t),\mu_0(t))$ 
      \vspace{5 pt}\Statex \begin{center}
      \textbf{Initialization}\end{center}
          \Statex \emph{Pre-Projection}
          \State $[A_0(t),B_0(t)]\gets\eqref{eq:LinDyn}$, with $(x(t),u(t))=\eta_0$
          \State $[Q_r(t),R_r(t),\Pi_r]\gets\eqref{eq:cost_arbitrary}$
          \State Compute $K_r(t)$ \Comment{B.i.t. \eqref{eq:DRE_reg}}
          
          \vspace{5 pt}\Statex \emph{Initial Trajectory and Starting Cost}
          \State $\xi_k\gets\mathcal P(\eta_0)$\Comment{F.i.t. \eqref{eq:simp_Proj} }
      \State $g_k\gets\eqref{eq:cost}$
      \Statex \begin{center}
          \textbf{Main Loop}
      \end{center}
      \Statex $Dg = -10\,\textrm{tol}$
      \While{$-Dg\geq \textrm{tol}$} 
      \Statex \emph{Compute Tangent Space and Linear Terms}
      \State $[A_k(t),B_k(t)]\gets\eqref{eq:LinDyn}$
      \State $[q_k(t),r_k(t),\pi_k]\gets\eqref{eq:LinCost}$\vspace{5 pt}
          \Statex \emph{Compute Regulator and Adjoint State}
        \State $[Q_r(t),R_r(t),\Pi_r]\gets\eqref{eq:cost_arbitrary}$
      \State Compute $K_r(t)$ \Comment{B.i.t. \eqref{eq:DRE_reg}}
      \State Compute $\chi_k(t)$ \Comment{B.i.t. \eqref{eq:adjoint}}\vspace{5 pt}
      \Statex \emph{Perform (quasi-)Newton Step}
      \State \textbf{try}
      \State $\quad~\!$ $[Q_k(t),S_k(t),R_k(t),\Pi_k]\gets$\eqref{eq:Newton}
      \State $\quad~\!$ Compute $[K_o(t),v_o(t)]$ \Comment{B.i.t. 
      \eqref{eq:DRE}}
      \State \textbf{catch} \eqref{eq:DRE} failed to converge \textbf{do}
      \State $\quad~\!$ $[Q_k(t),S_k(t),R_k(t),\Pi_k]\gets$\eqref{eq:quasi_Newton}
      \State $\quad~\!$ Compute $[K_o(t),v_o(t)]$ \Comment{B.i.t. \eqref{eq:DRE}}
      \State \textbf{end try}
      \State Compute $\zeta_k(t)$\Comment{F.i.t. \eqref{eq:update}}\vspace{5 pt}
      \Statex \emph{Compute Cost Gradient}
      \State $Dg_k\gets\pi_kz_k(T)+ \int_0^Tq_k(t)z_k(t)+r_k(t)\nu_k(t)dt$\vspace{5 pt}
      \Statex \emph{Armijo Step}
      \State $\sigma\gets\min(\,1\,,\,0.6\| x_0\|/\max(\|z_k(t)\|)\,)$
      \State $\xi_{k+1}\gets\mathcal P(\xi_k(t)+\sigma\zeta_k(t))$\Comment{F.i.t. \eqref{eq:simp_Proj} }
      \State $g_{k+1}\gets\eqref{eq:cost}$
      \While{$g_{k+1}>g_k-0.4\,  Dg_k$}
      \State $\sigma\gets0.7\,\sigma$
      \State $\xi_{k+1}\gets\mathcal P(\xi_k(t)+\sigma\zeta_k(t))$\Comment{F.i.t. \eqref{eq:simp_Proj} }
      \State $g_{k+1}\gets\eqref{eq:cost}$
      \EndWhile\vspace{5 pt}
      \Statex \emph{Proceed to Next Iteration}
  \State $\xi_k\gets\xi_{k+1}$
      \State $g_k\gets g_{k+1}$
  \EndWhile\Statex \begin{center}
          \textbf{Output}\end{center}
      \Statex \emph{Solution}
\State \textbf{return} $[x(t),u(t)]=\xi_k$
\Statex \hrulefill
\Statex F.i.t. = Integrate forward in time
\Statex B.i.t. = Integrate backward in time
\end{algorithmic}
\end{algorithm}

\end{document}